\documentclass[12pt,draftclsnofoot,onecolumn]{IEEEtran}

\usepackage{graphicx}
\usepackage{amsmath,bbm}
\usepackage{amsfonts}
\usepackage{cleveref}
\usepackage{verbatim}
\usepackage{comment}
\usepackage{cases}
\usepackage[inline]{enumitem}
\crefrangelabelformat{}{#3#1#4 to~#5#2#6}
\crefrangeformat{equation}{(#3#1#4) to~(#5#2#6)}
\usepackage{algorithm}
\usepackage{algpseudocode}
\usepackage{color}
\usepackage{multirow}
\usepackage[noadjust]{cite}
\usepackage{subcaption}
\newtheorem{prop}{Property} 
\usepackage{tagging}
\usepackage[colorinlistoftodos,textsize=tiny,textwidth=22mm]{todonotes}

\def\syn{synchronization }
\IEEEoverridecommandlockouts
\begin{document}

\title{Optimal  state replication in stateful data planes} 


\author{\IEEEauthorblockN{
Abubakar Siddique Muqaddas,
German Sviridov, 
Paolo Giaccone, 
Andrea Bianco}
\IEEEauthorblockA{Politecnico di Torino, Torino, Italy}

}

\maketitle


\begin{abstract}

In SDN stateful data planes, switches can execute algorithms to process traffic based on local states. This approach permits to offload decisions from the controller to the switches, thus reducing the latency when reacting to network events.
We consider distributed network applications that process traffic at each switch based on local replicas of network-wide states. Replicating a state across multiple switches poses many challenges, because the number of state replicas and their placement affects both the data traffic distribution and the amount of synchronization traffic among the replicas.  

In this paper, we formulate the optimal placement problem for replicated states, taking into account the data traffic routing, to ensure that traffic flows are properly managed by network applications, and the synchronization traffic between replicas, to ensure state coherence. Due to the high complexity required to find the optimal solution, we also propose an approximated algorithm to scale to large network instances. We numerically show that this algorithm, despite its simplicity,  well approximates the optimal solution. We also show the beneficial effects of state replication with respect to the single-replica scenario, so far considered in the literature. Finally, we provide an asymptotic analysis to find the optimal number of replicas. 

\end{abstract}
\begin{IEEEkeywords}
Software Defined Networking (SDN), Stateful data planes, State replication.
\end{IEEEkeywords}


\section{Introduction}

In recent years a major shift of paradigm has been observed in the field of SDN with the introduction of stateful data planes, which address the performance limitations of a complete centralization of the control plane in a canonical SDN architecture, as highlighted in~\cite{ref1,yeganeh2013scalability}. Indeed, stateful switches, as described for example in~\cite{bosshart2013forwarding,bianchi2016open}, can be programmed to execute user-defined code during packet processing, operating on local state variables stored in persistent memories.
Thus, stateful data planes provide an additional level of programmability with respect to canonical SDN, whose data plane is instead  stateless, according to the original paradigm.
Indeed, stateful switches can take local decisions without relying on the intervention of an SDN controller~\cite{7997297}. This fact has many beneficial effects. First, it greatly improves the reactivity of network applications by reducing the communication and latency overhead due to the interaction with the controller. Second, it reduces the computational burden of the controller to sustain the correct network behavior~\cite{he2015measuring}. Finally, 
 the availability of state variables enables the definition of new fine-grained networking applications~\cite{kim2015band}, as decisions can now be taken on a per-packet basis, contrary to the per-flow basis of canonical SDN. 

The availability of local state variables (simply denoted as ``states'' in the remainder of the paper) and the capability to run local programs (i.e., finite state machines) based on such states open a new perspective, since distributed algorithms can be devised to run in the switches across the network. This permits to extend the scalability of many network applications, thanks to the distributed nature of the approach. 

Differently from previous works, we focus on the specific scenario in which the network application runs locally in stateful switches on the basis of some non-local states. Indeed, for applications implementing network-wide policies, the value of a state may be ``global'' across multiple switches, each switch holding  a local replica of the state. Recent works, as~\cite{netsoft,arxiv}, have shown the practical feasibility of this approach by leveraging available programmable data planes, such as P4~\cite{bosshart2013forwarding} and Open Packet Processor (OPP)~\cite{bianchi2016open}. 



When a given state is replicated across multiple switches, two fundamental and coupled questions must be addressed: i) How many replicas are needed? ii) In which switches should replicas be placed? 
To find an optimal solution, several issues should be addressed. First, all traffic flows must traverse at least one switch that holds a state affecting (or affected by) the flow. However, routing a flow possibly not along its shortest path increases the data traffic load on the network. Thus, from the point of view of the data traffic, it would be convenient to increase the number of replicas until at least one replica is present along the shortest path of each flow. At the same time, adopting replicas comes at the cost of keeping the replicas synchronized. This requires the interaction between  switches holding the replicas, thus introducing a synchronization traffic, which increases with the number of replicas. This traffic affects the overall offered load on the network. Thus, from this perspective, it would be convenient to reduce the number of replicas as much as possible. In summary, the optimal selection of the number of replicas and their location depends on the tradeoff between the load introduced in the network by data and synchronization traffic.

In this paper, we address all the above mentioned questions and provide the following contributions:
\begin{itemize}
\item we propose the optimal state replication problem and formalize it as an ILP problem, that minimizes the overall (i.e., data plus synchronization) traffic load; 
\item to cope with the limited scalability of the ILP solver, we propose an approximation algorithm, denoted as \textsc{PlaceMultiReplicas} (PMR), able to solve large instances of the problem;
\item we numerically evaluate the performance of PMR and show that it well approximates the optimal solution, at least for small instances of the problem. Furthermore, we show that adding few replicas in a network can largely improve the performance with respect to the single-replica scenario; 
\item we analytically find the optimal number of replicas for unwrapped Manhattan network topologies and characterize its asymptotic behavior; we show that the formula obtained for large networks can be used also for small instances of the network.
\end{itemize}

The remainder of the paper is organized as follows. In Sec.~\ref{sec:repl}, we describe the state replication problem. In Sec.~\ref{sec:opt}, we present the ILP formalization of the optimal state replication problem. In Sec.~\ref{sec:algo}, we propose the PMR algorithm. In Sec.~\ref{sec:perf} we show the numerical results for the state placement problem. In Sec.~\ref{sec:asymp}, we present the asymptotic analysis of the  optimal number of replicas in a network. In Sec.~\ref{sec:statefulrelated} we discuss the related works. Finally, we draw our conclusions in Sec.~\ref{sec:statefulconc}.

\section{State replication in stateful SDN}\label{sec:repl}



Following the increasing need for highly dynamic network services and policies, the introduction of programmable data planes enables traffic processing policies to be offloaded directly into the switches. 
New frameworks to embed user-defined network policies to the stateful switches have been proposed~\cite{snap16,mcclurg2016event}. In this paper, we  consider SNAP~\cite{snap16} as a reference framework, even if our proposed approach is general and relevant to any programming abstractions for stateful data planes. 

SNAP introduces a one-big-switch (OBS) model as a network abstraction: the whole network (switches and links) is seen as a single ``big'' switch with a given set of input and output ports, corresponding to the end hosts, and an aggregate list of available resources for traffic processing. Due to the way the OBS abstraction is defined, flow routing between hosts is described on the basis of I/O port pairs. 
When defining a network application, the programmer is exposed to the OBS abstraction, without any knowledge of the actual underlying composition of the network.
The network applications are decomposed by SNAP into an extension of forward decision diagram (xFFD) that incorporates also stateful processing elements available at switches. The placement of the single-replica state affects the application and network performance.
Indeed, the xFFD and the traffic matrix between the OBS ports are fed into the SNAP ILP (Integer Linear Programming) optimizer, which selects the switches where to place each state and the corresponding processing logic of the decomposed application. 
The order in which the traffic traverses the switches storing the states plays a fundamental role, as state dependencies must be preserved to correctly execute the xFDD of the original application. 
To guarantee the correct execution of a network application, all flows affected by or affecting a state must be routed across the switch storing it. Thus, the routing does not generally follow the shortest path between the input and output OBS port, and the SNAP solver jointly optimizes the placement of the states and the routing to minimize the total data traffic load in the network.



The main limitation of SNAP emerges from the fact that it permits only one replica for each state. This considerably restrains the flow routing, thus precluding a wide range of optimization techniques such as load balancing and traffic engineering. 
\subsection{State replication}\label{sec:repl2}

To cope with the above mentioned SNAP limitations, we consider a scenario in which states are replicated on stateful switches. We address the optimal placement of the replicas of each state, given the knowledge of the traffic demands and of the xFDD defining the network application. 

\begin{figure}[bt!]
	\centering
	\begin{subfigure}[b]{.4\linewidth}
		\centering
		\includegraphics[scale=0.2]{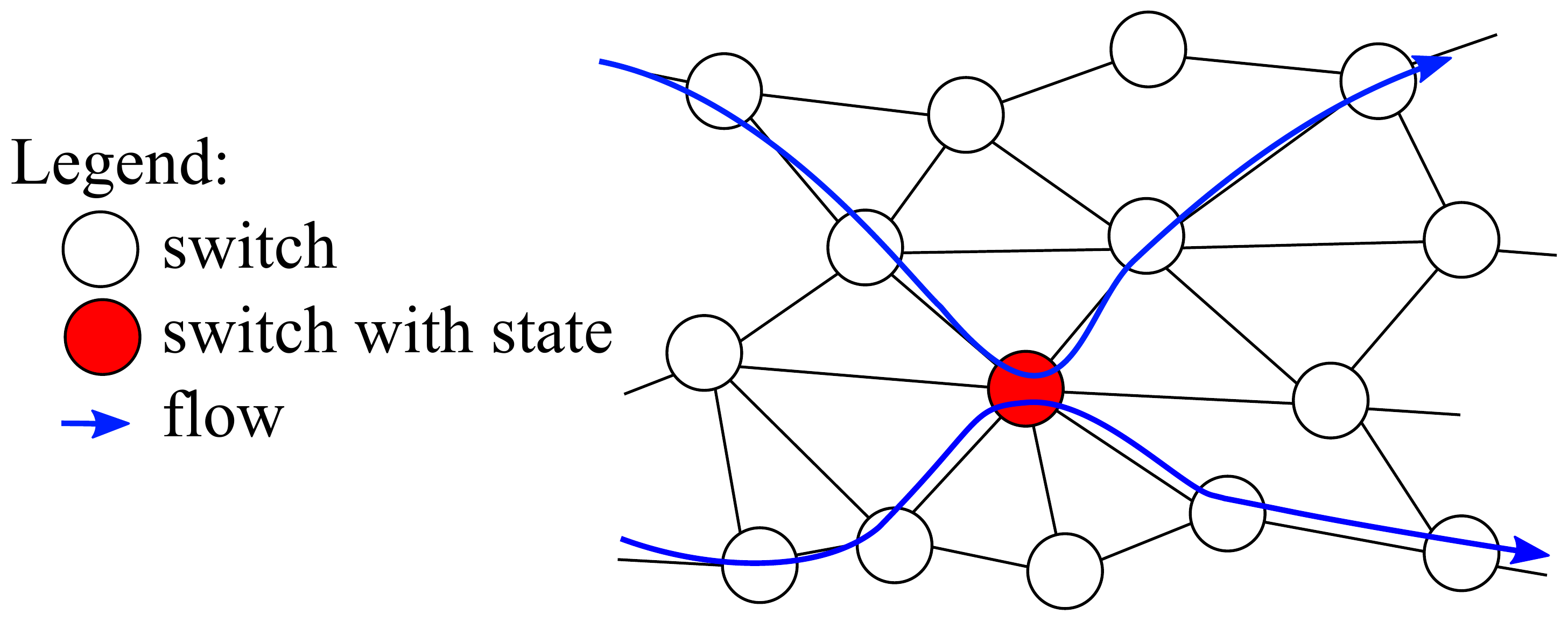} 
		\caption{Single-replica approach}
		\label{fig:snapfig}
	\end{subfigure}
	\begin{subfigure}[b]{.4\linewidth}
		\centering
		\includegraphics[scale=0.2]{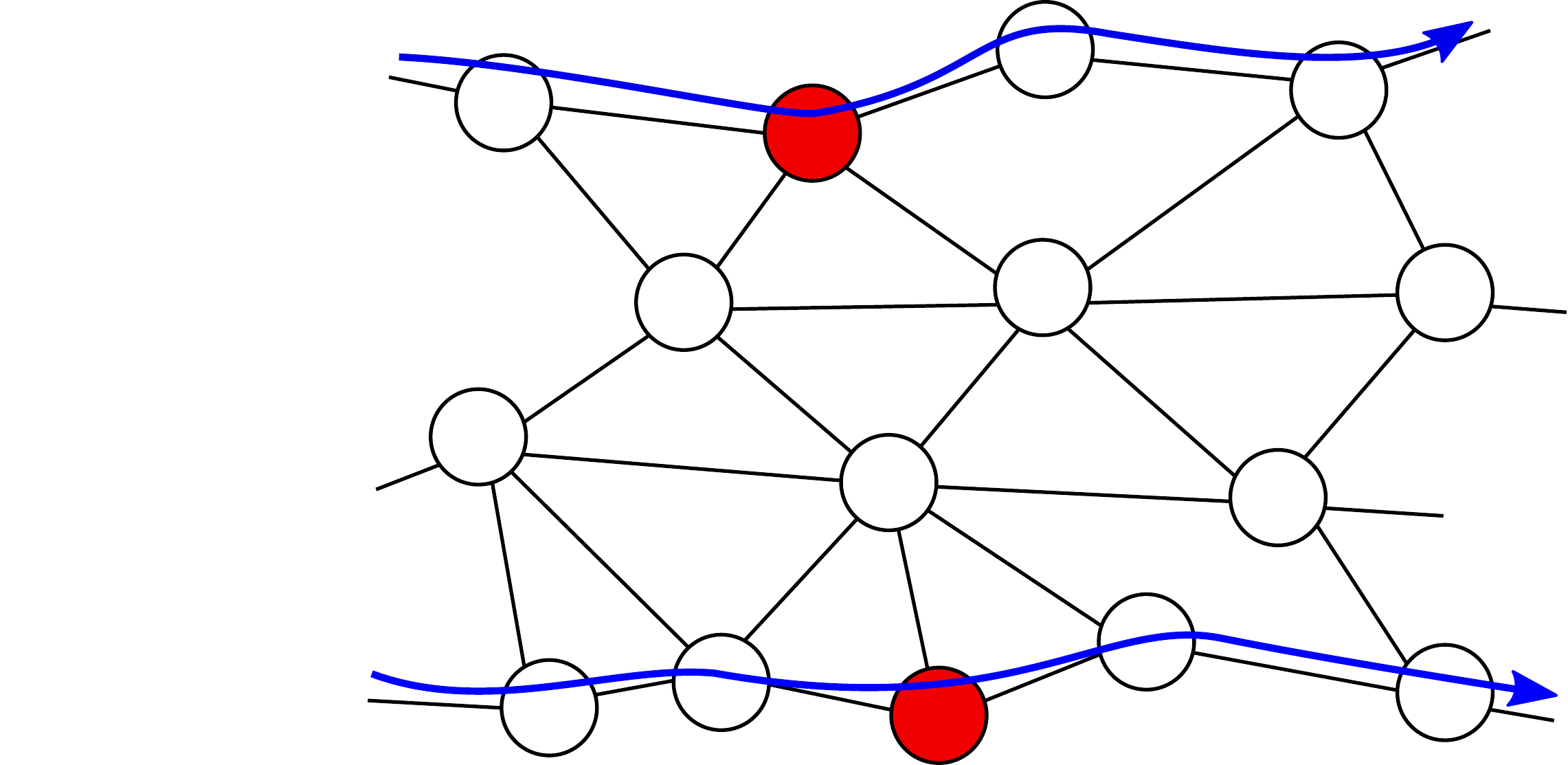} 
		\caption{Multi-replica approach}
		\label{fig:oursoln}
	\end{subfigure}
	\caption{Example of routing for single-replica (e.g., SNAP) and multi-replica state placement.}%
	\label{result1}%
\end{figure}

{As a toy example, consider a network-wide application that acts on a global counter (e.g., the total traffic entering/leaving the network), which is obviously affected by all flows in the network.
SNAP would place a single replica of the state associated with the global counter in a single  switch in the topology, likely into the switch in the most ``central'' position (i.e., with the highest betweenness centrality) in the network topology, as shown in Fig.~\ref{fig:snapfig}. As a consequence, all flows are forced to be routed through the single switch storing the state. Due to the ``hot-spot'' routing, the set of feasible solutions for the capacitated routing problem is significantly reduced.
Instead, replicating the global state on multiple switches would lead to a better network utilization, as shown in Fig.~\ref{fig:oursoln}, and to a much larger set of feasible routing solutions, with a beneficial effect on the maximum amount traffic that can be sustained in the network and/or on the experienced delays.}


The choice of an appropriate synchronization mechanism is crucial for network performance and for the implementation complexity of the replication scheme.
%
%
%
Notably, the CAP theorem~\cite{ref11} states that for a replication scheme, only two properties can be picked at the same time out of Consistency, Availability and Partition tolerance. Considering that network failures may occur, partition tolerance cannot be left out of the design of our replication algorithm, leaving us with the following, well-known, reference models:
\paragraph{Strong consistency}
A replication algorithm based on strong consistency privileges consistency over availability. This translates into strong guarantees that the same value of a state will be read across all replicas, at the cost of higher delays to access and update the states. The delay penalty is caused by the adopted protocol (e.g., Paxos~\cite{paxos}, Raft~\cite{raft}) requiring intensive interaction among the replicas whenever a read or write transaction is executed. Side effects of the  replication protocol are the  high overhead in terms of synchronization traffic and its high complexity, typically incompatible with the limited amount of hardware resources available at the switches. Furthermore, the latency due to the communication between replicas requires buffering packets at each switch while waiting for the outcome of the replication transaction. This further makes the scheme too complex to be adopted in practice in high speed networks.
\paragraph{Eventual consistency}
Replication schemes based on eventual consistency prioritize replicas availability over their consistency. This translates into low latencies during the execution of transactions at the cost of no guarantees on the consistency of the actual values of each replica. Most of eventual consistency algorithms are based on gossip protocols~\cite{birman2007promise, shapiro2011conflict, petersen1996bayou} which incur into small overhead in terms of  synchronization traffic. At the same time, due to the simplicity of the adopted communication protocols, these algorithms can be implemented in programmable switches.


Due to the implementation and performance issues highlighted for strong consistency schemes, we assume a replication scheme based on eventual consistency, according to which each replica generates a fixed amount of \syn traffic towards all the other replicas.
As shown in~\cite{netsoft}, this scheme can be implemented in current state-of-art programmable data plane and, in practice, maintains small errors among the values of the replicas.

\section{Optimal state replication problem}\label{sec:opt}

Given a network graph, the objective of the state replication problem is to identify the best set of nodes (i.e., switches) where to place the replicas of each state and to compute the optimal routing. Coherently with~\cite{snap16}, the nodes are selected to minimize the overall traffic in the network and to guarantee that all flows affecting (or affected by) a given state will traverse at least one state replica. Differently from~\cite{snap16}, the traffic in the network is composed not only of data traffic, but also of the traffic introduced by the synchronization protocol required to keep consistent the replicas of a given state. 

We propose an integer linear program (ILP) formalization, as in the original SNAP model~\cite{snap16}. The relevant notation is reported in Tab.~\ref{tab:milpNotationInput}. Our formalization takes the following input parameters:
%
\begin{itemize}
\item {\em Network.} Let $G=(V,E)$ be the network graph with $N$  nodes.  Let $c_{e}$ be the capacity of edge $e \in E$.
\item {\em Traffic flows.}
Let $\mathcal F$ be the set of all flows.
The traffic demands are assumed to be known in advance. In particular:
let  $\lambda_{f}$ be the demand of traffic flow $f \in \mathcal F$, being $f_s\in V$ and $f_d \neq f_s \in V$ respectively the source and the destination nodes of the flow. 
\item {\em State variables.}
Let $S$ be the set of all state variables. Let $S_{f}\subseteq S$ be the {\em ordered} sequence of state variables for flow $f\in\mathcal F$, obtained from the xFFD of the corresponding application.
\item {\em Maximum number of replicas.} Let $C_s$ be a given upper bound on the number of replicas  for a state variable $s$, chosen by the network designer. 
Note that the optimal number of replicas for state $s$, denoted by $\hat{C}_s$, will be computed while satisfying the constraint  $\hat{C}_s\leq C_s$. 
\end{itemize}

Let $H_{f}$ be the set of all possible sequences of state replicas for a flow $f$.
Consider a toy example in which a flow $f$ requires 3 state variables $\mathcal A$, $\mathcal B$, $\mathcal C$, i.e., $S_f=[\mathcal A,\mathcal B,\mathcal C]$. Each state has 2 replicas (denoted as ``1'' and ``2'').  Now $H_f=\{[1 1 1], [1 1 2], [1 2 1], [1 2 2], [2 1 1], [2 1 2], [2 2 1], [2 2 2]\}$, and, as example, the sequence $h=[1 2 1]$ implies that $f$ traverses replica~$1$ of state $\mathcal A$, then replica~$2$ of state $\mathcal B$, and finally replica~$1$ of state $\mathcal C$.
Let $h_s$  be the replica of state variable $s$ in sequence $h\in H_f$. For the above example with $h=[1 2 1]$, $h_{\mathcal A}=1$, $h_{\mathcal B}=2$ and $h_{\mathcal C}=1$.


The output of the solver is described as follows, and the relevant notation is reported in Tab.~\ref{tab:milpNotationOutput}:
\begin{itemize}
  \item {\em Placement of the replicas of each state.} 
  Let $P_{scn}$ be a binary variable  equal to 1 iff
   replica  $c$ of state $s$ is stored at node~$n$.
Note that the optimization problem might place multiple replicas on the same node, but this would correspond to a single instance of the state. Thus, the optimal number of distinct replicas $\hat{C}_s$ of state $s$ across the whole network can be computed as follows\footnote{Let $\mathbbm{1}_{\{A\}}$ be the indicator function of $A$, equal to 1 iff condition $A$ is true.}:
 \[
 \hat{C}_s=\sum_{n\in V} \mathbbm{1}_{\Big\{\displaystyle\sum_{c\leq C_s} P_{scn}>0\Big\}} 
 \]
  \item {\em Data traffic routing.} 
  Let  $R_{fhe}$ be a binary variable equal to 1 iff flow $f$ traverses the sequence of state replicas $h$ on edge $e$. The set of such variables describes the complete routing of all flows in the network, taking also into account the constraint for the required sequence of traversed replicas. To avoid out-of-sequence problems, we do not permit flow splitting between different sequences of replicas.
%
  \item {\em Synchronization traffic routing.}
  Let $\hat{R}_{snme}$ be a binary variable equal to 1 iff there are replicas of the state variable $s$ on nodes $n$ and $m$ and the flow from node $n$ to node $m$ traverses edge $e$. This set of variables describes the routing of the synchronization traffic between different replicas of the same state.
  Let $\hat{\lambda}_s$ be the traffic generated by each state replica to update each other single replica of the same state.
 \iftagged{journal}{{\color{red}
   Notably, this traffic is independent of the actual data traffic on the network, {\color{red}coherently with the adopted implementation, as discussed in Sec.~\ref{sec:impl}.?}}}
%
\end{itemize} 


\begin{table}[tb!]
  \footnotesize 
  \centering
  \caption{ Input variables
  \label{tab:milpNotationInput}}
  \begin{tabular}{c c l l }
    \hline
    \textbf{Context} & 
    \textbf{Variable} & \multicolumn{1}{c}{\textbf{Description}} & \multicolumn{1}{c }{\textbf{Range}} \\ \hline 
 \multirow{4}{*}{Network definition} &  {$V$} & {set of all nodes} & $\{1,\ldots,N\}$ \\     
 & $N$ & number of nodes (i.e., $|V|$)  & $\mathbb N$ \\ 
    & {$E$} & {set of all edges} \\ 
    & $c_{e}$ & capacity of edge $e\in E$ & $> 0$ \\ \hline
   \multirow{4}{*}{Flow definition}&
    $\mathcal F$ & set of all the flows\\
    & $\lambda_f$ & traffic demand for flow $f\in\mathcal F$ & $> 0$\\ 
    &$f_s$ & source node for flow $f\in\mathcal F$ & $1,\ldots, N$ \\ 
    &$f_d$ & destination node for flow $f\in\mathcal F$ & $1,\ldots, N$ \\ \hline
    \multirow{4}{*}{State definition}&
      $S$ & set of all state variables &  \\ 
    & $C_s$ & max number of replicas for state $s$ & {$\ge 1$} \\
         & $S_{f}$ & sequence of state variables for flow $f\in\mathcal F$ & $\subseteq S$ \\
     & \multirow{2}{*}{$\hat{\lambda}_s$} & synchronization traffic between & \multirow{2}{*}{$>0$} \\
   & & any pair of replicas for state $s\in S$ & \\
      \hline 

      \end{tabular}
\end{table}


\begin{table}[tb!]
	\footnotesize 
	\centering
	\caption{Output variables
		\label{tab:milpNotationOutput}}
		\begin{tabular}{clll}
			\hline
			\textbf{Context} & \multicolumn{1}{c}{\textbf{Variable}} & \multicolumn{1}{c}{\textbf{Description}} & \multicolumn{1}{c}{\textbf{Range}} \\ \hline 
		 Data  traffic &	\multirow{2}{*}{$R_{fhe}$} & 1 iff flow $f$ along sequence of replicas $h$ & \multirow{2}{*}{Binary} \\ 
	routing		&& traverses edge $e$  \\
			\hline
Synchronization&			\multirow{2}{*}{$\hat{R}_{snme}$} &1 iff synchronization traffic from node $n$  to  node $m$& \multirow{2}{*}{Binary} \\
  traffic  routing      &   & containing replicas of  state variable $s$ traverses edge $e$  \\ \hline
Replica &			\multirow{2}{*}{$P_{scn}$} & 1 iff replica $c$ of state $s$ is stored & \multirow{2}{*}{Binary}  \\ 
placement		&	& in node $n$ \\
			\hline
		\end{tabular}
\end{table}

\begin{table}[tb!]
	\footnotesize 
	\centering
	\caption{Auxiliary Variables
		\label{tab:aux}}
		\begin{tabular}{cll}
			\hline
			\multicolumn{1}{c}{\textbf{Variable}} & \multicolumn{1}{c}{\textbf{Description}} & \multicolumn{1}{c}{\textbf{Range}} \\ \hline 
			   $E_I(n)$ & set of edges entering node $n\in V$ & $\subseteq E$ \\
    $E_O(n)$ & set of edges leaving node $n\in V$ & $\subseteq E$ \\
    $E(n)$ & set of all edges incident to node $n\in V$ & $\subseteq E$ \\
     $H_{f}$ & set of all sequences of replicas for flow $f\in\mathcal F$  & - \\ 
    $h_s$ & replica id of state $s$ for flow $f \in\mathcal F$  in sequence $h\in H_f$ & $1,\ldots, C_s$
    \\        
            $P_{fsce}$ & 1 iff flow $f$ on edge $e$ has passed replica $c$ of state $s$  & Binary \\
			$X_{fh}$ & 1 iff flow $f$ is assigned $h \in H_{f}$ & Binary \\ 
         $U_{sn}$ &1 iff at least one replica of state variable $s$ is on node $n$ & {Binary} \\
			\multirow{1}{*}{$Y_{snme}$} &{1 iff $\hat{R}_{snme} > 0 $} & \multirow{1}{*}{Binary} \\ \hline

		\end{tabular}
\end{table}
Finally, Tab.~\ref{tab:aux} reports the list of auxiliary variables adopted in the ILP formalization.
	
In the optimal state replication problem, the total traffic in the whole network is minimized: 
\begin{equation}\label{eq:objTotalTraffic}
\min \sum_{e \in E}\sum_{f\in\mathcal F} \sum_{h \in H_{f}} R_{fhe}\lambda_{f} + \sum_{e \in E}\sum_{s \in S}\sum_{n \in V} \sum_{\substack{m \in V\\ n \ne m}} \hat{R}_{snme}\hat{\lambda}_{s}
\end{equation}
The first term represents the total data traffic in the network. It is obtained by summing all the traffic due to $f$ on all the possible sequences of state replicas and on all of the edges. Instead, the second term is the synchronization traffic between replicas of the same state, summed across all states and edges in the graph. 
{Notably, \eqref{eq:objTotalTraffic} is similar to  the objective function used by the SNAP framework in~\cite{snap16}, but with the introduction of the second term that takes into account the synchronization traffic, not included in SNAP.}

As an alternative, the objective function could be modified to minimize the maximum congestion on a link, obtained by summing data and synchronization traffic, as follows:
\begin{equation}\label{eq:objCongestion}
\min \max_{e\in E} \Big( \sum_{f\in \mathcal F} \sum_{h \in H_{f}} R_{fhe}\lambda_{f} + \\\sum_{s \in S}\sum_{n \in V} \sum_{\substack{m \in V\\ n \ne m}} \hat{R}_{snme} \hat{\lambda}_s\Big)
\end{equation}
and could be easily integrated in the following formalization, using well-known ILP modeling techniques.

%

\subsection{Constraints in the optimization problem}

We now discuss all the constraints considered in the ILP model. In some cases, we will get products of binary variables, but the corresponding constraint can be easily linearized according to well-known techniques.

\subsubsection{Data routing constraints}  Constraints  \eqref{eq:milp1}-\eqref{eq:milp4} are similar to the constraints for the classic multi-commodity flow problem. However, our modification consists of assigning a commodity for each sequence $h\in H_f$ of state variable replicas directly at the source of the flow $f$, to model the sequence of states required by each flow.



We introduce an auxiliary variable, which is an indicator function $X_{fh}$ equal to 1 if sequence $h \in H_{f}$ is assigned to flow $f\in\mathcal F$. 
\begin{equation}\label{eq:milp7}
	X_{fh} = \sum_{e\in E_O(f_s)}R_{fhe} - \sum_{e\in E_I(f_s)} R_{fhe} 
\end{equation}
Indeed, whenever a particular sequence $h$ is adopted, similar to \eqref{eq:milp1}, the net outgoing data traffic from source $f_s$ is 1. Notably, the second term considers the special case in which the flow is re-entering (and leaving) $f_s$ in the path to reach the state and then the destination.  We now force only one sequence $h$ to be assigned to flow $f$. $\forall f\in\mathcal{F}$:
\begin{equation}\label{eq:milp1}
	\sum_{h\in H_{f}} X_{fh}=1 
	\end{equation}
A similar constraint is defined for flow $f$'s destination $f_d$, but now the net incoming flow should be 1. $\forall f\in\mathcal F $:
\begin{equation}\label{eq:milp2}
\sum_{h\in H_{f}} \Big( \sum_{e\in E_I(f_d)} R_{fhe} - \sum_{e\in E_O(f_d)} R_{fhe}\Big)=1 
\end{equation}
The sum of all the data and synchronization traffic passing an edge must not exceed its capacity. $\forall e \in E$:
\begin{equation}
  \sum_{f\in\mathcal F}  \sum_{h\in H_{f}}R_{fhe}\lambda_{f}
  + \\\sum_{s \in S}\sum_{n \in V} \sum_{\substack{n^* \in V\\ n \ne n^*}} \hat{R}_{snn^*e} \hat{\lambda}_s
  \le c_{e} \label{eq:milp3}
  \end{equation}
Finally, the standard flow conservation condition must be satisfied at any node. $\forall h \in H_{f}, \forall f\in\mathcal F$:
\begin{equation}
	\sum_{e\in E_I(n)}R_{fhe}=\sum_{e\in E_O(n)}R_{fhe} \quad \forall n \in V \setminus \{f_s,f_d\}\label{eq:milp4}
\end{equation}

\subsubsection{Placement constraints}


Each replica can only be placed at one switch. 
$\forall s \in S, ~\forall c \leq C_s$:
\begin{equation}\label{eq:milp6}
	\sum_{n\in V}P_{scn} = 1 \quad 
\end{equation}

We now constrain the flows to be routed through the corresponding states, i.e., all flows dependent on a state must traverse the node where the replica of such state is located (except at source $f_s$ and destination $f_d$). $\forall n \in V \setminus \{f_s,f_d\}, \forall f\in \mathcal{F}, \forall h\in H_f , \forall s \in S_f$:
\begin{equation}\label{eq:milp8}
\sum_{e\in E_I(n)}R_{fhe} \ge P_{sh_s n} + X_{fh} - 1 
\end{equation}
Indeed, if a particular sequence $h$ is adopted for $f$, then \eqref{eq:milp8} becomes
\(
\sum_{e\in E_I(n)}R_{fhe} \ge P_{sh_s n}
\)
and in the case the node contains a replica $h_s$ of the state $s$, then 
\(
\sum_{e\in E_I(n)}R_{fhe} \ge 1
\), 
which forces at least one $R_{fhe}$ variable to be one on the incoming edges to $e$. Otherwise, if the sequence $h$ is not adopted for $f$, then \eqref{eq:milp8} becomes a useless bound.


We now define a variable that tracks the fact that a flow has already traversed a particular state along its path. For a flow $f$ traversing a replica $h_s$ of state $s$,   we define $P_{fsh_se}=0$ for all edges along the path before entering the node with replica $h_s$ of $s$, and $P_{fsh_se}=1$ for all edges on the path after $h_s$.
%
%
%
It is initialized to zero for all unused replica sequences $h$.
$\forall f \in \mathcal{F}, \forall s\in S_f, \forall h \in H_f, \forall e \in E$:
\begin{equation}\label{eq:milp10}
P_{fsh_se} \le R_{fhe} \quad 
\end{equation}
To model the fact that $P_{fsh_se}$ changes from 0 to 1 whenever the flow  leaves a node where the state is stored, we set:
$\forall f\in~\mathcal{F}, \forall s\in S_f, \forall h \in H_f, \forall e \in E,  \forall n \in V \setminus \{f_s,f_d\}$:
\begin{equation}\label{eq:milp11}
	P_{sh_sn}X_{fh}+\sum_{e\in E_I(n)}P_{fsh_se}=\sum_{e\in E_O(n)}P_{fsh_se} 
\end{equation}
Indeed, only when $P_{sh_sn}X_{fh}=1$ (i.e., node $n$ has replica $h_s$ and $f$ exploits $h$ including it), the net flow of $P_{fsh_se}$ entering $n$ is 0 and the corresponding one leaving $n$ is 1.

We now impose that the data flow reaches the destination $f_d$ after having traversed all the states required in $h$, i.e.\ 
$P_{fsh_s e}=1$ for one edge entering $f_d$.
 $\forall f \in \mathcal{F}, \forall s\in S_f, \forall h \in H_f$:
\begin{equation}\label{eq:milp12}
P_{sh_sf_d}X_{fh}+\sum_{e\in E_I(f_d) }P_{fsh_se}=X_{fh} 
\end{equation} 
 
 So far, the constraints \eqref{eq:milp10}-\eqref{eq:milp12} force the flows to pass through all the required state variables, but not necessarily in sequence.  
We model here the correct sequence of traversed states, if the flow $f$ has to cross $h_s \in H_f$ of $s$, followed by replica $h_{s'} \in H_f$ of $s'$.
$\forall f \in \mathcal{F}, \forall s,s'\in S_f, \forall h \in H_f, \forall n \in V$
\begin{equation}\label{eq:milp13}
P_{sh_sn} + \sum_{e\in E_I(n) }P_{fsh_se} \ge P_{s'h_{s'}n} + X_{fh}  - 1 
\end{equation}
Indeed, if either  flow $f$ has been assigned sequence $h$, i.e., $X_{fh}=1$, or replica $h_{s'} \in H_f$ exists at node $n$,
or  replica $h_s \in H_f$ does not exist at node $n$, 
then \eqref{eq:milp13} becomes 
\(
\sum_{e\in E_I(n) }P_{fsh_se} \ge 1
\).
This forces $P_{fsh_se}$ to be 1 before entering node $n$, which means that the flow must have traversed $h_s$ before entering the node containing $h_{s'}$. This ensures that the flow traverses the correct sequence of states as dictated by $h$.

Constraint \eqref{eq:milp14} ensures that if flow has traversed state variable replica $h_s$ on edge $e$, i.e., $P_{fs'h_{s'}e}=1$, then it must have already crossed state variable replica $h_s$, which ensures $P_{fsh_se}=1$.
$\forall f \in \mathcal{F}, \forall s, s'\in S_f, \forall h \in H_f, e \in E$:
\begin{equation}\label{eq:milp14}
P_{fsh_se} \ge P_{fs'h_{s'}e} 
\end{equation}
%
%
%

\subsubsection{State synchronization}

State synchronization implies the generation of  synchronization traffic between any pair of replicas of the same state. Thanks to the routing variable $\hat{R}_{snme}$, we can model the traffic between any pair of nodes $n$ and $m$ containing replicas of the state variable $s$ and consider its contribution in the total traffic, as in~\eqref{eq:objTotalTraffic} and \eqref{eq:objCongestion}, and in the constraint~\eqref{eq:milp3} regarding the edge capacity. 

In the optimization model, multiple replicas of the state variable can be hosted on the same node $n$. Hence, to track that there is at least one replica at node $n$, we define the variable $U_{sn}$ in \eqref{eq:sync1}. $\forall c \in C_s,\: \forall s \in S,\: \forall n \in V$:
\begin{equation}\label{eq:sync1}
U_{sn} \ge P_{scn} 
\end{equation}
For the synchronization traffic from node $n$ to node $m$, the routing variable $\hat{R}_{snme}$ is treated as a commodity from node $n$ such that $U_{sn}=1$ to node $m$ such that $U_{sm}=1$.  
We constrain the routing to ensure the standard flow conservation equation at the intermediate node. 

We define a new intermediate variable $Y_{snme}$, set to 1 iff $\hat{R}_{snme} > 0$. This is ensured using the big-M method~\cite{Luenberger:2015:LNP:2843008} as in~\eqref{eq:sync2} where M is sufficiently larger than $\hat{R}_{snme}$. $\forall s \in S,\: \forall n \in V,\: \forall m \ne n \in V,\: \forall e \in E$
\begin{equation}\label{eq:sync2}
0 \le -\hat{R}_{snme} + MY_{snme} \le M - 1
\end{equation}
To fix a large enough value for $M$, assume $\hat{R}_{snme}=1$, $\forall e \in E_O(n)$, then $Y_{smne}=1$ from~\eqref{eq:sync2}. In this case, for the condition $M \ge \hat{R}_{snme}$ to be true, $M$ must be equal to or greater than the maximum degree of $G$: 
\begin{equation}\label{eq:bigm}
M \ge \Delta_G
\end{equation}
with $\Delta_G = \max_{n \in V}|E_O(n)|$.

We require the egress synchronization flow from a state replica containing node to use only one outgoing edge. This can be done by exploiting $Y_{snme}$ as in~\eqref{eq:sync3}. $\forall s \in S,\: \forall n \in V,\: \forall m \ne n \in V$:
\begin{equation}\label{eq:sync3}
\sum_{e \in E_{O(n)}} Y_{snme} \le 1
\end{equation}

The following constraints~\eqref{eq:sync4}-\eqref{eq:sync7} model the  multi-commodity flow problem for the synchronization traffic. 
Specifically, constraints~\eqref{eq:sync4} and~\eqref{eq:sync5} are for the originating synchronization flow from the source node $n$ and the sink flow in the destination node $m$ containing the state replicas respectively. $\forall s \in S,\: \forall n \in V,\: \forall m \ne n \in V$: 
\begin{equation}\label{eq:sync4}
\sum_{e \in E_{O(n)}} Y_{snme} \ge U_{sn}
\end{equation}
\begin{equation}\label{eq:sync5}
\sum_{e \in E_{I(m)}} Y_{snme} \ge U_{sm}
\end{equation}

Instead, constraints~\eqref{eq:sync6}-\eqref{eq:sync7} are for the flow conservation at intermediate nodes. $\forall s \in S,\: \forall n \in V,\: \forall m \ne n \in V$:
\begin{equation}\label{eq:sync6}
\sum_{e \in E_{O(n)}} Y_{snme} \le \sum_{e \in E_{I(n)}} Y_{snme} + U_{sn} \le 1
\end{equation}
\begin{equation}\label{eq:sync7}
\sum_{e \in E_{I(n)}} Y_{snme} \le \sum_{e \in E_{O(n)}} Y_{snme} + U_{sm} \le 1
\end{equation}

\tagged{journal}{ {

BUG: THE SYNCH TRAFFIC MUST BE ONLY FROM DIFFERENT replicas OF 

We consider that the the synchronization traffic generated by each state variable replica  $c$ to every other replica $g$ of each state variable $s$ with the variable: $\hat{R}_{scge}$, $\forall s \in S,\:\forall c \ne g \leq C_s$. 

Assume the case where state replica $c$ of variable $s$ needs to synchronize with state replica $g$ of the same state variable. At the node $n$ where $P_{scn}=1$ i.e. the source, there must be $\hat{R}_{scge}=1$ on one outgoing edge, whereas for all incoming edges, $\hat{R}_{scge}$ must be equal to 0. Vice versa, at the node $n$ where $P_{sgn}=1$ i.e. the destination, there must be $\hat{R}_{scge}=1$ on one incoming edge, whereas for all outgoing edges, $\hat{R}_{scge}$ must be equal to 0. For all intermediate nodes, $\hat{R}_{scge}$ on incoming edges must be equal to $\hat{R}_{scge}$ on all outgoing edges. These mentioned conditions can be shown to be equivalent to the constraints \Crefrange{eq:milp15}{eq:milp18} $\forall s \in S,\:\forall c \ne g \leq C_s,\:\forall n\in V$.
\begin{gather}
\sum_{e \in E_O(n)} R_{scge} \le \sum_{e \in E_I(n)} R_{scge} + P_{scn} \le 1\label{eq:milp15}\\
\sum_{e \in E_I(n)} R_{scge} \le \sum_{e \in E_O(n)} R_{scge} + P_{sgn} \le 1\label{eq:milp16}\\
\sum_{e \in E_O(n)} R_{scge} \ge P_{scn}\label{eq:milp17}\\
\sum_{e \in E_I(n)} R_{scge} \ge P_{sgn} \label{eq:milp18}
\end{gather}

The amount of the state synchronization traffic can also be proportional to the amount of data traffic entering the switch containing the state variable. This assumption modifies the objective function \eqref{eq:objTotalTraffic} as in \eqref{eq:objTotalTrafficProp}.

\begin{multline}\label{eq:objTotalTrafficProp}
\min \quad \sum_{f \in \mathcal{F}} \sum_{h \in H_f} \sum_{e \in E}R_{fhe}\lambda_f+ \\ \sum_{s \in S} \hat{\lambda}_s \sum_{c \leq C_s} \sum_{f \in \mathcal{F}} \sum_{e' \in E_I(n)} \sum_{h \in H_f} R_{fhe'} P_{scn} \sum_{g \ne c \leq C_s} \sum_{e \in E}  R_{scge}
\end{multline}
where the product of decision variables can be linearized.
}}

\iftagged{journal}{
In the worst case,  in the proposed ILP formalization the number of variables grows as 
$O(N^3) $
and the number of constraints grows as
$O(N^4)$.
}

\tagged{journal}{{ RECHECK ALL VALUES}}

%

\subsection{Computational complexity}

The complexity to solve an ILP model is $O(2^{2^{k_v+2}}k_c)$~\cite{Megiddo84}, %
where $k_v$ is the number of variables and $k_c$ is the number of constraints.
As a worst case, assume that all flows $f \in \mathcal F$ require to traverse all state variables $s \in S$, where each $s \in S$ has $C$ replicas. In this case, it can be shown that 
\(k_v=O(\max(N^2C^{|S|},|S|N^4))\) and 
\(k_c=O(\max(N|S|C^{|S|},|S|N^4))\).
In a simple scenario when only one state variable required by all the flows, 
$k_v=O(N^4)$ and $k_c=O(N^4)$. Thus, the final complexity is lower bounded by 
\(
O(2^{2^{N^4+2}}N^4)
\).
%
%
Clearly, the presented ILP formalization does not scale for large instances of the problem. This advocates the design of approximation algorithms to solve the optimal replication problem in real scenarios, as addressed in the following section.


\section{Approximation algorithm for  single state replication}\label{sec:algo}

We address specifically the problem of state replication for a single state variable. To address the limited scalability of the ILP solver, we propose \textsc{PlaceMultiReplicas} (PMR) algorithm which is computationally scalable and will be shown in Sec.~\ref{sec:perf} to approximate well the optimal solution obtained by the ILP solver for small problem instances. 

The pseudocode of PMR is given in Algorithm~\ref{algo:placemultiplecopies}. It takes as input the network graph $G$, the state variable $s$ and the maximum number of replicas $C_s$ of $s$ and the set of flows $\mathcal F$ requiring~$s$. As output, the algorithm returns: the routing variables of the data flows $R_{fhe}$ and of the state synchronization flows $\hat{R}_{smne}$ and the replicas placement variables $P_{scn}$. The algorithm works through 3 phases:
 \begin{itemize}
 \item {\em Phase 1.} The network graph $G$ is partitioned into $C_s$ clusters, in order to minimize the maximum distance among the elements within a cluster. This allows to distribute the replicas across the whole network in a balanced way, exploiting the spatial diversity offered by each cluster. 
 \item {\em Phase 2.} In each cluster, a replica is placed in the ``most central''
 node, i.e., the one with the highest betweenness centrality,
 in order to minimize the data traffic for each flow.
 \item {\em Phase 3.} The position of each replica is perturbed at random using a local search  to improve the solution with respect to one obtained in the previous two phases.
 \end{itemize}
%
 
 \begin{algorithm}[!tb]
	\scriptsize
	\caption{PlaceMultiReplicas (PMR)}\label{algo:placemultiplecopies}
	\begin{algorithmic}[1]
		\Procedure{$[\{R_{fhe}\},\{\hat{R}_{smne}\},\{P_{scn}\}]$ = PlaceMultiReplicas}{$G$, $s$, $C_s$, $\mathcal F$} \label{algo1:init}
        	\State $R_{fhe}=0, \forall f\in\mathcal F, h\in H_f, \forall e\in E$\Comment{Init routing} \label{algo1:initS}
            \State $\hat{R}_{smne}=0, \forall c,g \ne c\leq C_{s}, \forall e\in E$\Comment{Init state sync} \label{algo1:init3} 
			\State $P_{scn}=0, \forall c \leq C_s, \forall n\in V$\Comment{Init state $s$ location} \label{algo1:init2} \label{algo1:initE}
			\State $\{G_c\} \gets$ \textsc{ComputePartitions($G,C_s,$)} \label{algo2:partition} \Comment{\textbf{Phase 1:}  Graph partitions $\{{G_c}\}$}
			\For{$c \leq C_s$} \label{algo2:bestmetric1}	\Comment{\textbf{Phase 2:} Replica placement}			
            	\State $n' \gets$ \textsc{NodeWithHighestBC($G_c$) } \Comment{Find best candidate in  partition $G_c$}
           		\State $P_{scn'} = 1$ \Comment{Store the state replica location}
			\EndFor \label{algo2:bestmetric2}
			\State $T_{\min} = \infty$ \Comment{Init minimum traffic}
			\For{$I$ iteration} \label{algo2:startLS} \Comment{\textbf{Phase 3:} Local search}
            	\State [$T',\{R_{fhe}^{\prime}\},\{\hat{R}_{smne}^{\prime}\}] \gets$ \textsc{RouteFlows($\mathcal F,\{P_{scn}\}$)} \label{algo2:returnTocurrent} \Comment{Route flows through the replicas}
				\If{$T' < T_{\min}$} \Comment{Check if the traffic is smaller} \label{algo1:condition}
					\State $T_{\min} = T'$ \Comment{Store current best solution} \label{algo1:update1}
					\State
					$R_{fhe} = R_{fhe}^{\prime}$ 
					$\hat{R}_{smne} = \hat{R}_{smne}^{\prime}$,
					$P'_{scn}=P_{scn}$, $\forall f\in \mathcal F$, $\forall h\in H_f$, $\forall c, g\neq c\leq C_s$, $\forall e\in E$, $\forall n\in V$ \label{algo1:update2}
					 
				\EndIf
                \State $\{P'_{scn}\}\gets $ \textsc{PerturbReplicaLocation($\{P_{scn}\}$)} \label{algo2:perturb} \Comment{Change existing location of state replicas}
               
			\EndFor \label{algo2:endLS} \\
            \Return $[\{R_{fhe}\}, \{\hat{R}_{smne}\},\{P_{scn}\}]$
		\EndProcedure
	{  
		\item[]
		\Procedure{[$T_\text{current},R_{fce}^{\prime},\hat{R}_{smne}^{\prime}]=$  RouteFlows}{$\mathcal F,P_{scn}$}
			\State $T_\text{current} = 0$ \label{algo2:tcurrent} \Comment{Init total traffic}
			\For{$f \in \mathcal{F}$} \label{algo2:routingflows1} \Comment{For each flow}
				\State $\text{minDist} = \infty$ \Comment{Init minimum distance}
				\State $c_{b} \gets \text{null}$ \Comment{Init best replica for current flow} \label{algo1:bestCopyInit}
                \State $\mathcal{P}_{best} \gets \text{null}$  \Comment{Path with minimum length for $f_s \to n_c \to f_d$} \label{algo1:bestPathInit1}
				\For{$c \in C_s$} \Comment{For all state replicas} \label{algo1:eachCopyInit}
                    \State $\mathcal{P} = \textsc{ShortestPath}(f_s,n_c) \cup \textsc{ShortestPath}(n_c,f_d)$
					\If{$\mathcal{P}.\text{length} < \text{minDist}$} \label{algo1:ifMinLen}
						\State $\text{minDist} = \mathcal{P}.\text{length}$ \Comment{Update minimum distance}
						\State $\mathcal{P}_{best} \gets \mathcal{P}$ \Comment{Store path with minimum length} 
                        \State $c_{b} \gets c$ \Comment{Store best replica for this flow}
					\EndIf
				\EndFor \label{algo1:eachCopyEnd}
				\For {$e\in \mathcal{P}_{best}$} \Comment{For each edge in the minimum length path} \label{algo1:bestPathInit2}
                 \State $R_{fc_{b}e}^{\prime} = R_{fc_{b}e}^{\prime}+\lambda_f$ \Comment{Store the routing}
                 \State $T_\text{current} = T_\text{current} + \lambda_f$ \Comment{Store the traffic value}
                \EndFor \label{algo1:bestPathEnd}
			\EndFor \label{algo2:routingflows2}           
			\For{$c \in C_s$} \label{algo2:routingsync1} \Comment{For each c\textsuperscript{th} replica of state variable $s$}
				\For{$g \ne c \in C_s$} \Comment{For each g\textsuperscript{th} replica of state variable $s$}
					\State $\mathcal{P}_{cg} \gets \textsc{ShortestPath}(n_c,n_g)$ \Comment{Shortest path from $n_c \to n_g$}
                    \For {$e\in \mathcal{P}_{cg}$} \Comment{For each edge in the path $n_c \to n_g$}
                 		\State $\hat{R}_{smne} = \hat{R}_{smne}+\alpha$ \Comment{Store the state sync flow} \label{algo1:updatingStateSync}
                 		\State $T_\text{current} = T_\text{current} + \alpha$ \Comment{Update total traffic} \label{algo1:updatingTcurrent}
                	\EndFor
				\EndFor 
			\EndFor \label{algo2:routingsync2}\\ 
			\Return $[T_\text{current},R_{fce}^{\prime},\hat{R}_{smne}^{\prime}]$
		\EndProcedure }
		\end{algorithmic}
	\end{algorithm}

Algorithm~\ref{algo:placemultiplecopies} comprises all the mentioned phases.
After having initialized the routing and the replica placement variables (lines~\ref{algo1:initS}-\ref{algo1:initE}), Phase 1 is executed in line~\ref{algo2:partition} by calling \textsc{ComputePartitions}. This method solves the $k$-means clustering problem~\cite{cluster} with $k=C_s$ using Lloyd's algorithm~\cite{voronoi} in which the node with the highest betweenness centrality is chosen as center of the partition.

As part of Phase 2 (lines~\ref{algo2:bestmetric1}-\ref{algo2:bestmetric2}), within each subgraph $G_c$  the node $n'$ with the highest betweenness centrality 
is assigned a state variable replica through \textsc{NodeWithHighestBC}.
As a reminder, betweenness centrality of a node $v$ is proportional to the number of shortest paths  crossing it. 

Lines~\ref{algo2:startLS} to~\ref{algo2:endLS} refer to a local search procedure with $I$ iterations. Within each iteration,  \textsc{RouteFlows} is used to route flows through the location of the replicas identified in  Phase 2,  
following two sub-paths: one from the flow source node to the closest replica and one from this replica to the destination node.
The procedure works on the set of flows $\mathcal{F}$ and the location of state variables $P_{scn}$ and returns the routing variables for data flows $R_{fce}^{\prime}$ and for state synchronization $\hat{R}_{smne}^{\prime}$, and the corresponding total traffic $T'$ in the network.
Lines~\ref{algo2:routingflows1} to~\ref{algo2:routingflows2} route the data flows from their source $f_s$ to the destination $f_d$ while traversing the replica $c_b$ which has the minimum path length among all other replicas. For each flow, in lines~\ref{algo1:bestCopyInit} and~\ref{algo1:bestPathInit1}, the replica $c_b$ and the path $\mathcal{P}_{best}$ traversing it are initialized. Then for each replica (in lines~\ref{algo1:eachCopyInit}-\ref{algo1:eachCopyEnd}), first, the shortest path $f_s \to n_c \to f_d$ is computed. $n_c$ is the vertex for which $P_{scn}=1$. If the path length $\mathcal{P}$.length is less than the previous minimum minDist in line~\ref{algo1:ifMinLen}, then the current path $\mathcal{P}$ is stored as the best path $\mathcal{P}_{best}$ and the current replica $c$ as the best replica $c_b$. In lines~\ref{algo1:bestPathInit2}-\ref{algo1:bestPathEnd}, for each edge in $\mathcal{P}_{best}$, the routing as well as the traffic value is updated. Lines~\ref{algo2:routingsync1} to~\ref{algo2:routingsync2} generate flows from each state replica $c$ to all the other state replicas $g$ for state synchronization using the shortest path. This includes the synchronization flows $\hat{R}_{scge}$ being updated in line~\ref{algo1:updatingStateSync} for each edge in the path $\mathcal{P}_{cg}$ before updating the total traffic in line~\ref{algo1:updatingTcurrent}.
If $T'$ is less than the previous minimum, then the minimum traffic value and all the decision variables are updated (lines~\ref{algo1:update1}-\ref{algo1:update2}). In Phase 3  (line~\ref{algo2:perturb}), a local search procedure perturbs the existing state replica locations. This proceeds by randomly selecting one node where a replica is located and moving it to one of its neighbor nodes. This new solution is then compared with the current one (line~\ref{algo1:condition}) after having evaluated the corresponding routing and total traffic.
\clearpage
{\color{blue}
\iftagged{journal}{
\subsection{Graph partitioning} \label{sec:graphPartitioning}
We utilize a Voronoi-region based graph partitioning technique~\cite{voronoi} while placing state variable replicas in the network so that replicas are spatially distributed and are closer to the flows. The distance among the nodes in the graph is based on the shortest path i.e. the number of hops. The pseudocode is given in Algorithm~\ref{algo2:computePartitions}. {\color{red} Should I put a figure depicting graph partitioning?} The inputs consist of the graph $G$, the number of state variable replicas $C_s$ to be placed, and the number of allowed iterations $I_p$ while partitioning. As an output, the algorithm returns $\hat{G}$, which is the set of all subgraphs, where $G_c \in \hat{G}$.

The explanation of the algorithm is as follows. At first, the set of subgraphs $\hat{G}$ is initialized in line~\ref{algo2:initG}. If number of replicas $C_s$ is only 1, then there is only one partition and the original graph $G$ is stored as the solution in line~\ref{algo2:singleGraph}. Instead for multiple replicas, lines~\ref{algo2:startMultipleCopyS}-\ref{algo2:startMultipleCopyE} are used. A set $\mathcal{V}$ is initialized in line~\ref{algo2:initVRecur} which is used to store all explored solutions of state replica locations. Initially, locations of the state replicas are chosen at random in line~\ref{algo2:random} and stored in the current solution $V_c$. Lines~\ref{algo2:PartitionS}-\ref{algo2:PartitionE} constitute the main loop of the algorithm responsible for the partitioning of the graph. It stops till either state replica locations $V_c$ recur or the number of iterations run is $I_p$ in line~\ref{algo2:PartitionS}. A new \emph{affiliation} vector $A[n]$ is initialized in line~\ref{algo2:affInit}, which affiliates every vertex in the graph to the closest state replica location. The current state replica locations $V_c$ is stored in $\mathcal{V}$ in line~\ref{algo2:saveCurrentSoln}. In lines~\ref{algo2:AffS}-\ref{algo2:AffE}, all the vertices in the graph are affiliated to the closest state replica location. In lines~\ref{algo2:affLeaderS}-\ref{algo2:affLeaderE}, if the current solution $V_c$ contains the current vertex $n$, this signifies that current vertex is one of the state replica locations and it is affiliated with itself in line~\ref{algo2:affItself}. For the rest of the vertices $V \setminus V_c$, the algorithm continues. In lines~\ref{algo2:nclosestInit} and~\ref{algo2:minDistInit}, the closest state replica location and the distance to it are initialized respectively. In lines~\ref{algo2:computeClosestS}-\ref{algo2:computeClosestE}, the closest state replica $n_{closest}$ is identified. The shortest path to the state replica location $n_c$ is computed and its length is compared to the previous minimum in line~\ref{algo2:compareMinimum}. If it is smaller, then the minimum distance is updated in line~\ref{algo2:updateDistance} and $n_c$ is stored in $n_{closest}$ in line~\ref{algo2:updateClosest}. After this, the current vertex $n$ is then affiliated to the closest replica $n_{closest}$ in line~\ref{algo2:affClosest}. Using the information of the affiliation, the vertices having the affiliation with the same state replica location are added to one subgraph. This is done for all the state replica locations and subgraphs are generated in line~\ref{algo2:subGraph}. After this, the current state replica locations $V_c$ are discarded for new locations which are computed in lines~\ref{algo2:computeNewS}-\ref{algo2:computeNewE}. Within each subgraph $G_c$, the vertex with the maximum betweenness centrality value is chosen as one of the new state replica locations in line~\ref{algo2:betweenness}. Then this process is repeated $I_p$ times or till a state replica location solution recurs.

\iftagged{journal}{ {\color{blue}
\begin{algorithm}
	\scriptsize
	\caption{Graph partitioning algorithm}\label{algo2:computePartitions}
	\begin{algorithmic}[1]
	\Procedure{$\hat{G} =$ ComputePartitions}{$G$, $C_s$, $I_p$}
		\State $\hat{G} \gets \text{null}$ \Comment{Initialize set of subgraphs} \label{algo2:initG}
		\If{$C_s=1$}
			\State $\hat{G} = G$ \Comment{If only one replica, the original graph $G$ is the only subgraph} \label{algo2:singleGraph}
      	\Else \label{algo2:startMultipleCopyS}
			\State $\mathcal{V} \gets \text{null}$ \Comment{Initialize set of explored state replica locations} \label{algo2:initVRecur}
			\State $V_c \gets \textsc{PlaceRandom}(G,C_s)\: \forall c\in C_s$ \Comment{\parbox[t]{.3\linewidth}{Initial random placement of state replicas}} \label{algo2:random}
			\State $i=0$
        	\While{$(i <I_p \text{ AND } !\mathcal{V}.\text{contains}(V_c)) $} \Comment{\parbox[t]{.3\linewidth}{Loop $I_p$ times or till state replica locations recur}} \label{algo2:PartitionS} 
            	\State $A[n] \gets \text{null}\: \forall n \in V$  \Comment{Initialize vertex to state replica affiliation} \label{algo2:affInit}
				\State $\mathcal{V}[i] = V_c\: \forall c\in C_s$ \Comment{\parbox[t]{.5\linewidth}{Add current state replica location to list of explored solutions}} \label{algo2:saveCurrentSoln}
                \For{$n \in V$} \Comment{For all vertices in the graph} \label{algo2:AffS}
					\If{$V_c.\text{contains}(n)$} \label{algo2:affLeaderS}
                    	\State $A[n] =n$ \Comment{Leader in each partition affiliates with itself} \label{algo2:affItself}
                        \State \textbf{continue}
                    \EndIf \label{algo2:affLeaderE}
                    \State $n_{closest} \gets \text{null}$ \Comment{Initialize closest state replica location} \label{algo2:nclosestInit}
                    \State $\text{minDist} = \infty$ \Comment{Initialize min distance to state replica location} \label{algo2:minDistInit}
                    \For{$n_c \in V_c$} \Comment{For all leaders in each partition} \label{algo2:computeClosestS}
						\State $\mathcal{P} = \textsc{ShortestPath}(n,n_c)$ 
                        \If{$\mathcal{P}.\text{length} < \text{minDist}$} \label{algo2:compareMinimum}
                        	\State $\text{minDist} = \mathcal{P}.\text{length}$ \Comment{Update current min distance} \label{algo2:updateDistance}
                           	\State $n_\text{closest} = n_c$ \Comment{Update closest replica} \label{algo2:updateClosest}
                        \EndIf
                    \EndFor \label{algo2:computeClosestE}
                    \State $A[n] = n_\text{closest}$ \Comment{Affiliate $n$ with closest state replica location} \label{algo2:affClosest}
                \EndFor \label{algo2:AffE}
				\State $\hat{G} \gets \textsc{CreateSubGraph}(G,A[n])$ \Comment{\parbox[t]{.3\linewidth}{Use affiliation to create subgraphs}} \label{algo2:subGraph}
				\State $V_c \gets \text{null}\:\forall c\in C_s$ \Comment{Discard leaders}
				\For{$c \in C_s$} \Comment{Computing new leader in each partition} \label{algo2:computeNewS}
					\State $V_c \gets \max(\textsc{Betweenness}(G_c))$ \Comment{\parbox[t]{.3\linewidth}{vertex with highest betweenness is chosen as state replica location}} \label{algo2:betweenness}
				\EndFor \label{algo2:computeNewE}
				\State $i=i+1$
			\EndWhile \label{algo2:PartitionE}
		\EndIf \label{algo2:startMultipleCopyE} \\
		\Return $\hat{G}$			
      \EndProcedure
	\end{algorithmic}
\end{algorithm}
}}

\subsection{Local search}\label{sec:localSearch}
The graph partitioning method is helpful in minimizing the data traffic in the network by distributing state replicas across the network. However, we also want to target the minimization of the synchronization traffic. To achieve this purpose, we perform Tabu search as a local search technique after obtaining the state replica locations and the total traffic before line~\ref{algo2:perturb} in the \textsc{PlaceMultiReplicas} algorithm. The \textsc{PerturbCopyLocation} routine refers to the Tabu search technique. This consists of first randomly picking one of current state replica locations $n_s$ out of $V_c$. Then this state replica is shifted to one of the randomly picked neighbors of the vertex $n_s$. This changes the current state replica location solution, for which the flows are routed again and it is checked if the solution reduces the traffic value. If there is no improvement, then the solution is reverted to the previous best. Then this process is repeated till $I$ iterations or if shifting the state replica location to any neighbor node does not improve the solution as indicated by a flag $STOP$.
}

}



\section{Performance comparison} \label{sec:perf}

We evaluate the performance of PMR presented in Sec.~\ref{sec:algo}.
The local search in PMR runs with $I=1000$ iterations. 
In the case of small instances of the problem, we run an ILP solver, coded using IBM CPLEX optimizer~\cite{ibmcplex}, implementing the optimization model  in Sec.~\ref{sec:opt}. 
{Notably, whenever the number of replicas is set to 1, $\hat{\lambda}_s=0$ and the solver obtains a solution equivalent to the one achieved by SNAP.}
We compute the {\em approximation ratio}, i.e.,  the ratio between the total traffic obtained by PMR  and the optimal traffic obtained by the ILP solver.
%
%
%
We consider two standard topologies for the network graph: 
%
%
\begin{itemize}
	\item {\em Unwrapped Manhattan} is a $\sqrt{N}\times \sqrt{N} $  grid. 
    \item {\em Watts-Strogatz}~\cite{watts1998collective} adds a few long-range links to regular graph topologies to reduce the distances between pairs of nodes and emulate a small-world model.
    It is generated by taking a ring of $N$ nodes, where each node is connected to $k$ nearest neighbors. In each node, the edge connected to its nearest clockwise neighbor is disconnected with probability $p$ and connected to another node chosen uniformly at random over the entire ring. Thus, the final topology maintains the original average degree $k$ while being connected. In the following, we will use $p=0.1$ and $k=8$.
\end{itemize}


\iftagged{journal}{
\subsubsection{Traffic} \label{sec:traffic}
}

We utilize random traffic matrices with the number of flows equal to the number of nodes in the graph ($|\mathcal F|=N$) and with unity demands ($\lambda_f=1$).
The source-destination pairs for the flows were generated according to two models. In the case of {\em uniform traffic}, all the source nodes were associated to a random permutation of nodes as destination;  thus each node is source and destination of exactly one flow. In the case of {\em clustered uniform traffic}, we partitioned the nodes of the graph in half and generated a random permutation between the nodes of the same partition; thus all the flow are local within the same partition. All the results were obtained with 1000 different runs to get very small 95\% confidence intervals (in all cases within 4.2\% accuracy).
%

\subsection{Synchronization traffic and optimal number of replicas}

In Fig.~\ref{fig:syncVsNumCopies} we evaluate the effect of varying the number of replicas for state $s$ and of the \syn  rate $\hat{\lambda}_s$, through the optimal ILP solver. 
We consider a $4 \times 4$ Manhattan graph and set $C_s=7$.
As expected, when increasing the traffic required to synchronize  the replicas ($\hat{\lambda}_s$), the optimal number of replicas reduces, since the higher costs of \syn compensates the beneficial effect of multiple replicas on the data traffic. Instead the \syn traffic is almost constant, since, for smaller number of replicas, their relative distances grows, to ``cover'' a larger area of the network. 
%
As a term of comparison, we report the total traffic for one single replica allowed in the network, {equivalent to the solution obtained by SNAP}. 


\begin{figure}[!tb]
	\centering
	\includegraphics[width=8cm,trim={10 0 30 5},clip]{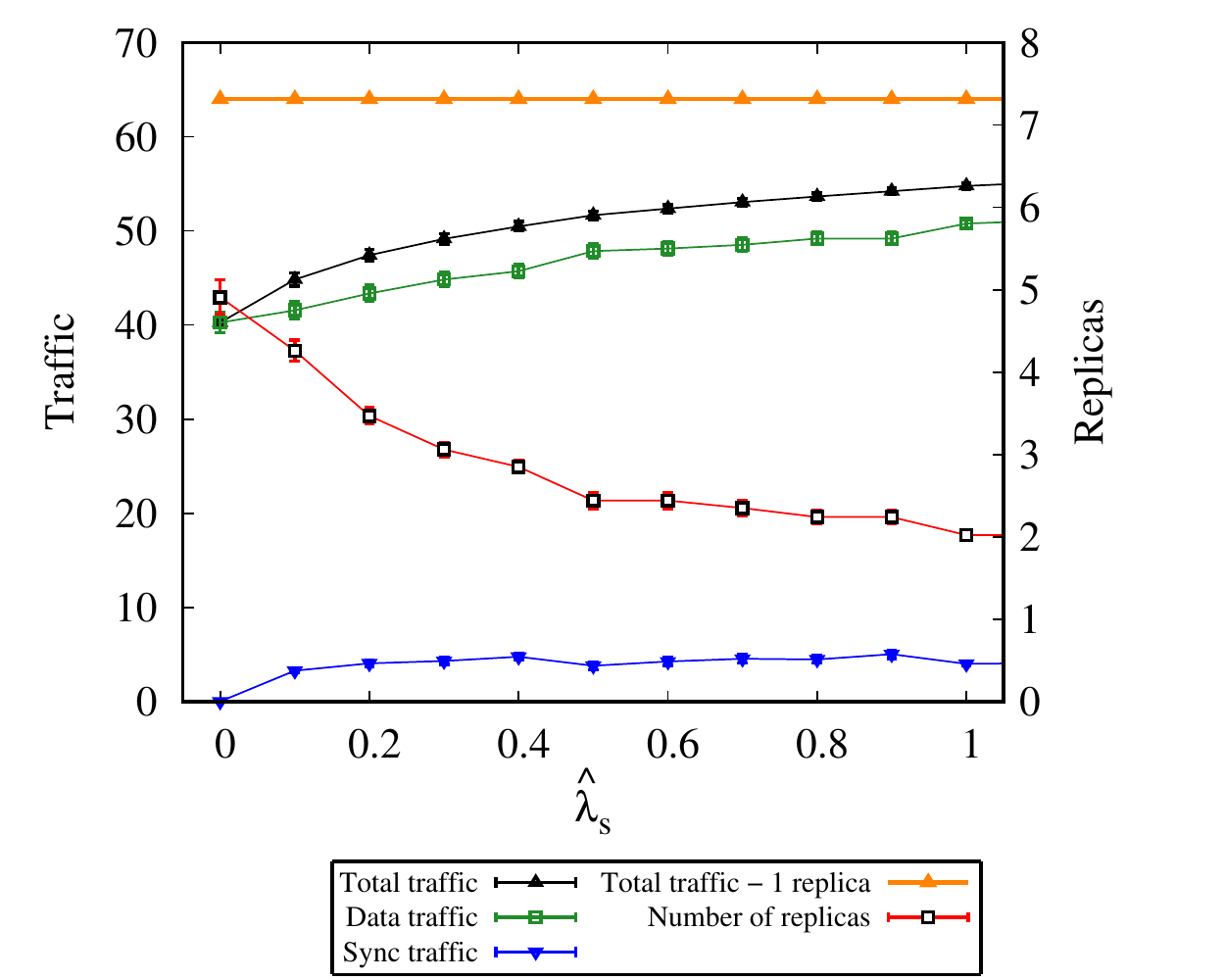}
	\caption{Optimal traffic and number of replicas in a  $4 \times 4$  Manhattan graph for uniform traffic, using the ILP solver. }\label{fig:syncVsNumCopies} 
\end{figure}

Fig.~\ref{fig:syncVsNumCopies2} extends Fig.~\ref{fig:syncVsNumCopies}  for larger values of $\hat{\lambda}_s$.  Due to the higher cost for synchronization, for $\hat{\lambda}_s \geq 6.1$, the optimal number of replicas becomes one, i.e., it is not anymore convenient to replicate states due to the high \syn cost {and the final solution is equivalent to the one achieved by SNAP.}

\begin{figure}[!tb]
	\centering
	\includegraphics[width=9cm,trim={5 0 25 0},clip]{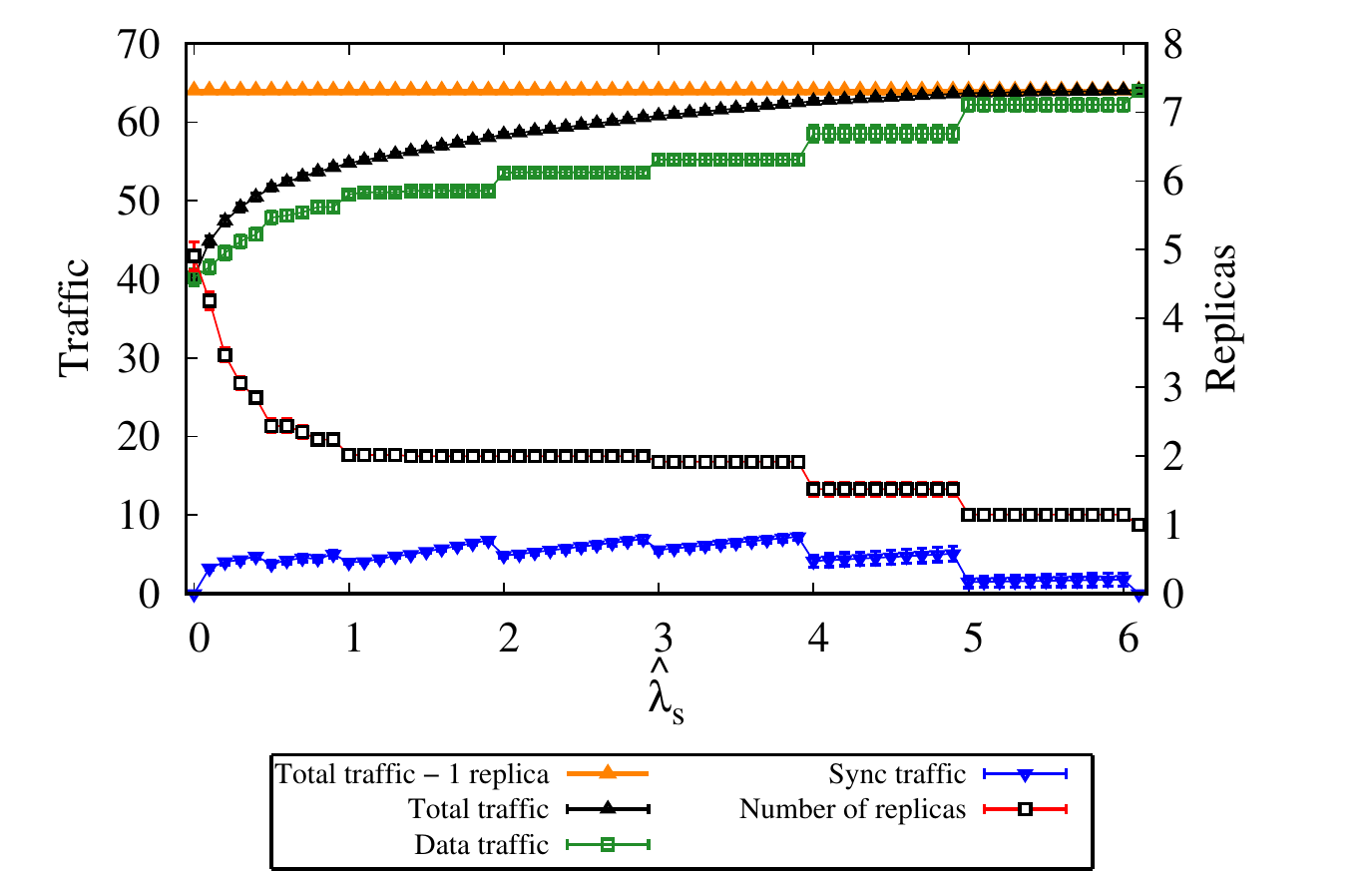}
	\caption{Optimal traffic and number of replicas in a  $4 \times 4$  Manhattan graph for uniform traffic, using the ILP solver, for large values of $\hat{\lambda}_s$. }\label{fig:syncVsNumCopies2}
\end{figure}

\subsection{Comparison of PMR with ILP}\label{sec:milp}


 Figs.~\ref{fig:approxManhattan}-\ref{fig:approxWattsStrogatz} show the 
 approximation ratio for different number of nodes $N$, of replicas $C_s$ and different values of $\hat{\lambda}_s$, under uniform traffic. The two graphs refer to Manhattan and Watts-Strogatz graphs, respectively. The approximation ratio in all cases is always $\le 1.15$, thus PMR approximates well the ILP solution. For larger graphs, we could not  provide the results as the ILP solver is not computationally feasible.
 


\begin{figure}[tb!]
	\centering
	\includegraphics[width=8cm]{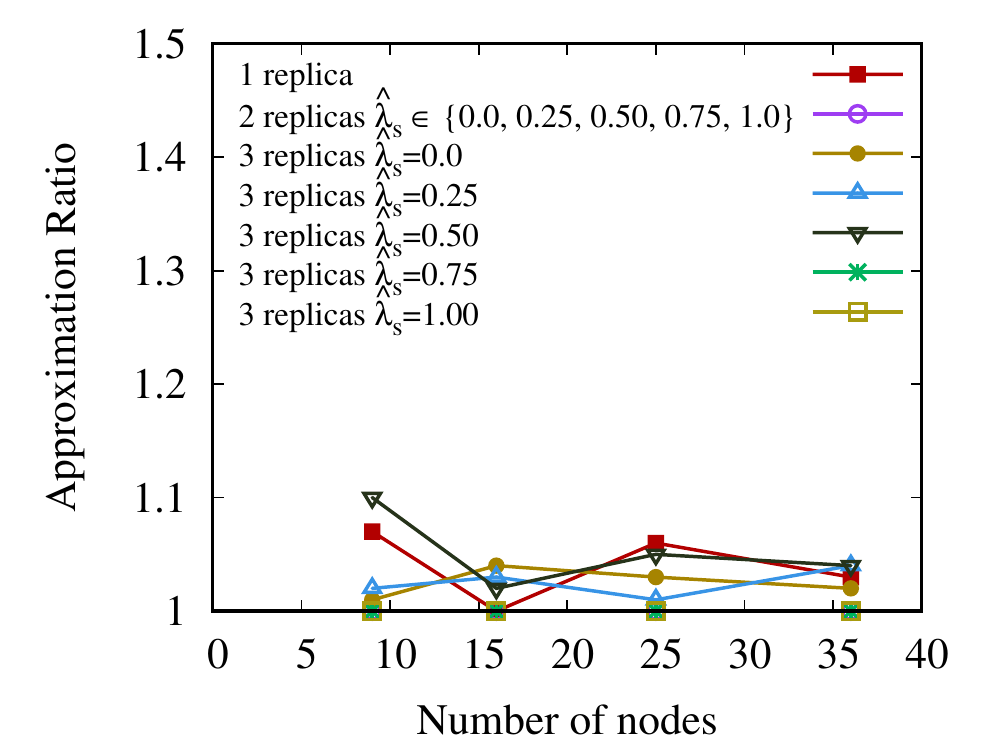}
	\caption{Approximation ratio of PMR in a Manhattan graph under uniform traffic. }\label{fig:approxManhattan}
\end{figure}


\begin{figure}[tb!]
	\centering
	\includegraphics[width=8cm]{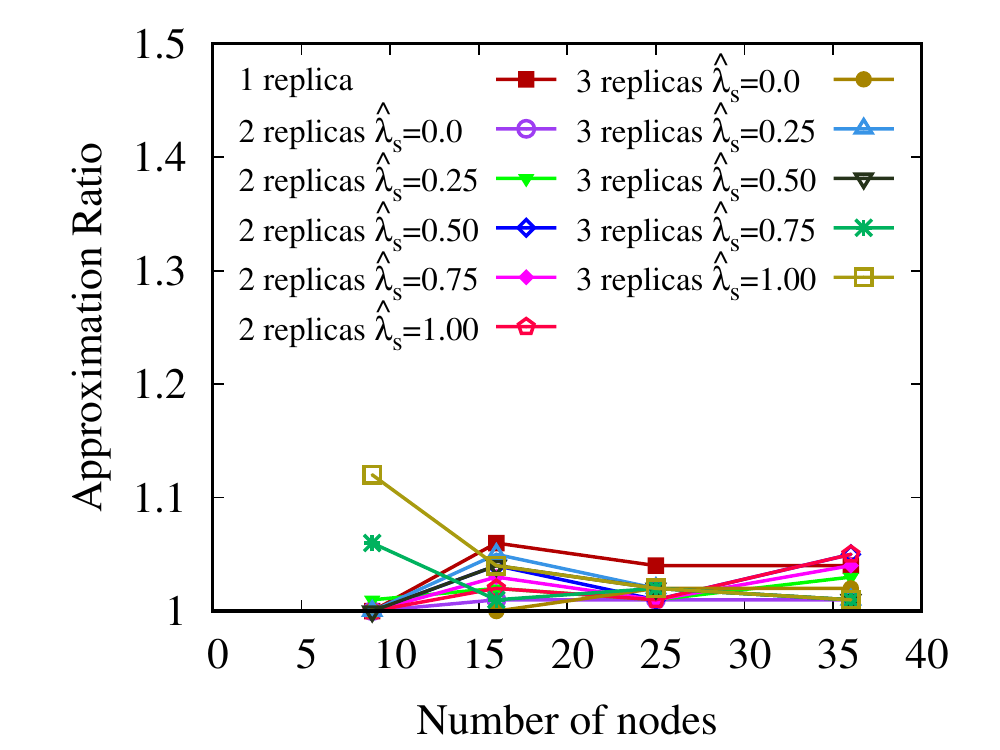}
	\caption{Approximation ratio of PMR in Watts-Strogatz graph under uniform traffic. }\label{fig:approxWattsStrogatz}
\end{figure}

\subsection{Number of replicas in large topologies}

For large topologies, we run just the PMR algorithm. Figs.~\ref{fig:stateRepManhattan}-\ref{fig:stateRepWattsStrogatz4} show the total traffic, normalized by the number of flows, for   Manhattan and Watts-Strogatz graphs, under clustered uniform traffic. We set $\hat{\lambda}_s=0.5$. 
For comparison, we also report the result of the traffic obtained by routing each flow from its source to its destination along the shortest path, obliviously of the placement of the state replicas; this provides a lower bound on the total traffic in the network obtained for the optimal solution of the ILP problem (which cannot be computed in this case).

As expected, the highest amount of traffic is given by the single-replica case, because of the longer path to reach the state location targeted  by all the flows. Now adding one replica provides a beneficial effect, since the spatial diversity of 2 replicas can be exploited to route the flows and minimize the total traffic. The gain is generally around 30\% for Manhattan graph and grows up to 20\% in Watts-Strogatz graph. If increasing again the number of replicas from 2 to 3, then the gain is very limited (around 5\%), since the higher spatial diversity is compensated by a higher \syn traffic. Thus, in general we can expect that allowing few replicas has a strong beneficial effects on the overall traffic with respect to the single-replica scenario.
%
		

\begin{figure}[tb!]
	\centering
	\includegraphics[width=8cm]{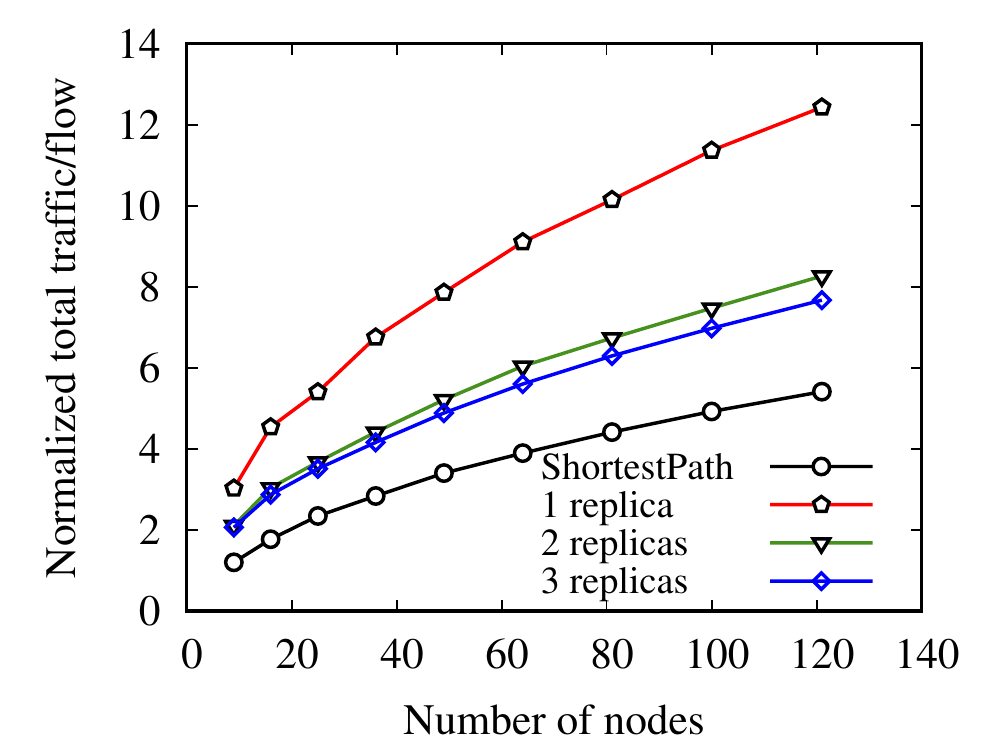}
	\caption{Performance of PMR in Manhattan  graph under clustered uniform traffic. }\label{fig:stateRepManhattan} 
\end{figure}

\iftagged{journal}{
\begin{figure}[tb!]
	\centering
	\includegraphics[width=6cm]{Figures/plot_wattsstrogatz_p_0_5_alpha_0_50_normalizedPerFlow_tabu.pdf}
	\caption{Performance of PMR in Watts-Strogatz graph \color{red} randperm(n), p=0.5, $TM=100$, $I_P=10$, $I=1000$, $|G|=1$}\label{fig:stateRepWattsStrogatz1}
\end{figure}

\begin{figure}[tb!]
	\centering
	\includegraphics[width=6cm]{Figures/{plot_wattsstrogatz_p_0_5_alpha_0_50_normalizedPerFlow_randperm_tabu}.pdf}
	\caption{Performance of PMR in Watts-Strogatz graph \color{red} - Randperm(n/2), p=0.5, $TM=100$, $I_P=10$, $I=1000$, $|G|=1$}\label{fig:stateRepWattsStrogatz2}
\end{figure}

\begin{figure}[tb!]
	\centering
	\includegraphics[width=6cm]{Figures/{plot_wattsstrogatz_p_0_alpha_0_50_normalizedPerFlow_randperm_tabu}.pdf}
	\caption{\color{red}Performance of \textsc{PlaceMultiReplicas} in Watts-Strogatz graph ($p=0$) under clustered uniform traffic  }\label{fig:stateRepWattsStrogatz3} 
\end{figure}
}

\begin{figure}[tb!]
	\centering
	\includegraphics[width=8cm]{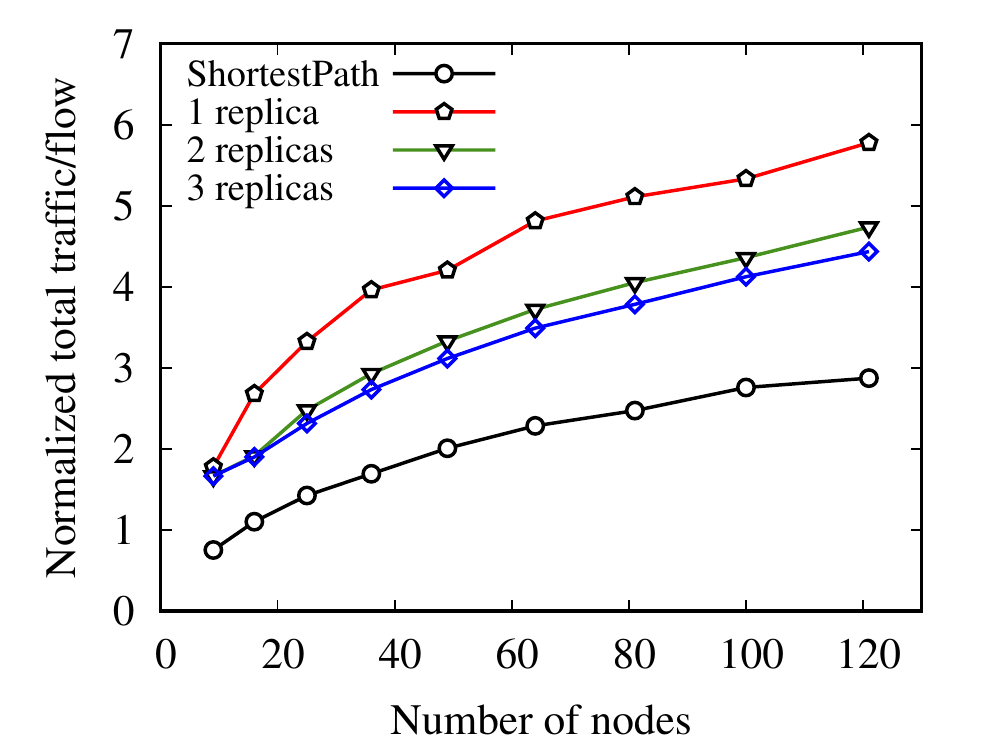}
	\caption{Performance of PMR in Watts-Strogatz graph under clustered uniform traffic.}\label{fig:stateRepWattsStrogatz4}
\end{figure}

{

\section{Asymptotic analysis for number of replicas} \label{sec:asymp}

We now present an asymptotic analysis, i.e., for very large network graphs, to estimate the optimal number of replicas. We will consider specifically an  unwrapped Manhattan topology since amenable to analytical modeling. Furthermore, for simplicity we assume a single state.

\subsection{Methodology}

We consider a unit square as shown in Fig.~\ref{fig:square}, representing the boundary of an unwrapped Manhattan topology containing $N$ nodes, with $N \to \infty$. Thus, any position within the unit square is associated to a network node, and any line within the unit square represents a routing path across a sequence of nodes in the original topology.

We now assume that the number of replicas $C$ is a perfect square, i.e. $\sqrt{C} \in \mathbb{N}$. The unit square is divided into individual $C$ squares, each of them of size  $1/\sqrt{C}\times 1/\sqrt{C}$ and with a center point $P^{ctr}_{c}$, where $c \in \{1,\ldots, C\}$ is an index identifying the square, as shown in Fig.~\ref{fig:square}. Here, $P^{ctr}_{c}$ denotes the location of the $c$-th state replica in the network. We now evaluate the  optimal number of replicas that  minimizes the total traffic in the topology. 

\begin{figure}[!tb]
	\centering
	\includegraphics[width=9cm]{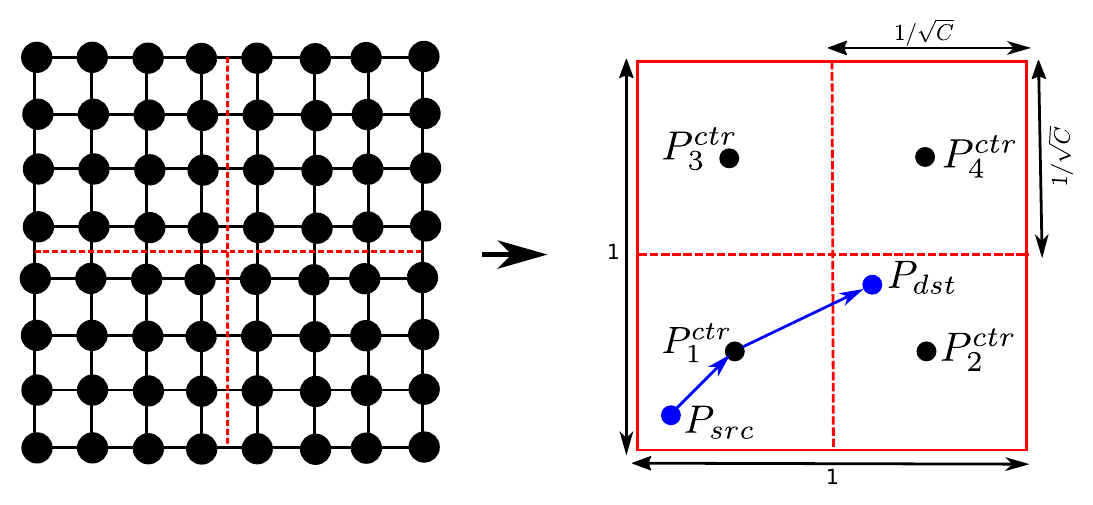}
	\caption{Unwrapped Manhattan topology (left) and its unit square representation with  4 replicas ($C=4$) (right).}
	\label{fig:square}
\end{figure}

The total traffic is composed of the data traffic and the synchronization traffic, coherently with the cost function in~\eqref{eq:objTotalTraffic}. Consider now a given flow $f\in\mathcal F$. We assume that the traffic demand $\lambda_f$ is routed in a straight line between two points in the square, since its approximates well the  step-wise stair-like routing in the original Manhattan topology, for $N\to\infty$. 
The total traffic generated by the flow is $\lambda_f h$ where $h$ is the corresponding distance of the routing path in terms of hops in the Manhattan topology.
 The following bound can be easily shown, relating the distance $d$ between two points in the unit square and the corresponding routing distance in terms of hops:
\begin{equation}\label{eq:bb}
d\sqrt{N}\leq h\leq d\sqrt{2} \sqrt{N}
\end{equation}

Now recall that a flow from a source node $P_{src}$ to a destination node $P_{dst}$ must traverse at least one replica $P^{ctr}_c$, as shown in Fig.~\ref{fig:square}, in order to affect (or being affected by) the state replica.

We start by evaluating the overall data traffic.
We assume uniform traffic between any pair of nodes in the original topology, with  a total number of flows equal to $|\mathcal F|=N$ and all flows with rate $\lambda_f$, coherently with Sec.~\ref{sec:perf}. 
Based on~\eqref{eq:bb}, we can define the average routing distance as:\begin{equation}
\hat{h}= \hat{d} \sqrt{N} \beta
\end{equation}
where $\beta$ is a constant value less than $\sqrt{2}$.
Thus, the overall data traffic generated in the network  can be computed as the total generated data traffic $\lambda_f  N$ times the average distance $\hat{h}$:
\begin{equation}\label{eq:tdata}
T_{data} = \lambda_f   \hat{d}_{data}N \sqrt{N}   \beta
\end{equation}
where $\hat{d}_{data}$ is the average total distance between two randomly generated points in the unit graph passing through the closest replica.

\begin{figure}[tb!]
	\centering
	\includegraphics[width=8cm]{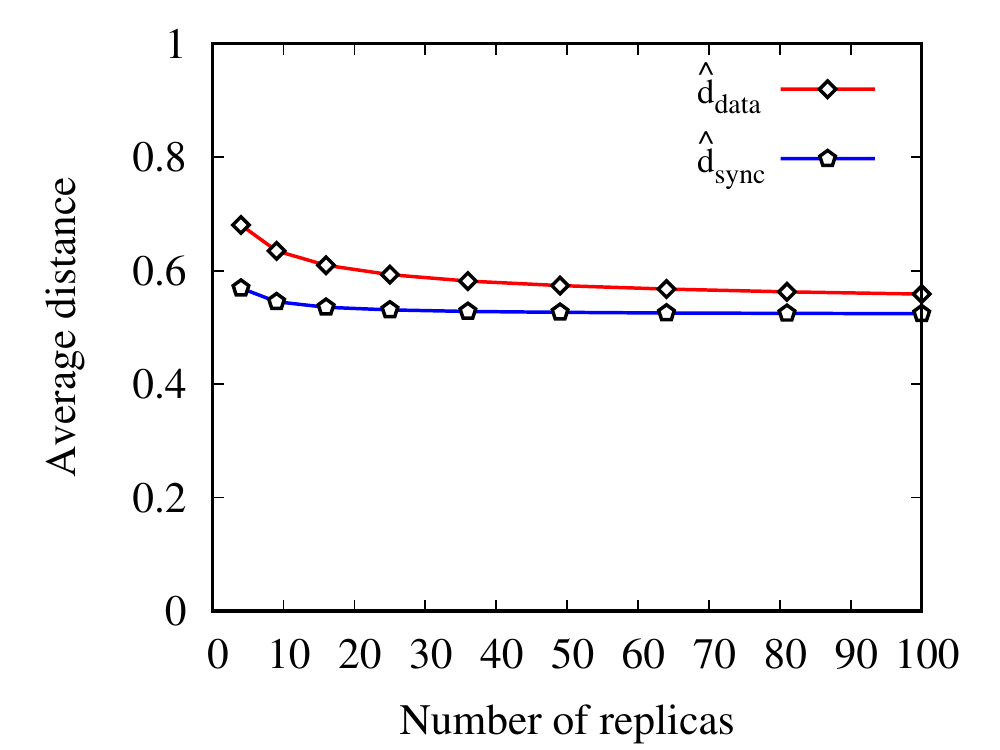}
	\caption{The total average distance $\hat{d}_{data}$ and $\hat{d}_{sync}$ in function of the number of replicas $C$ for a unit square.}
	\label{fig:avgdata}
\end{figure}

To evaluate $\hat{d}_{data}$,  we utilize a Monte Carlo method. We generate pairs of points with uniform random coordinates in the unit square, which are $P_{src}$ and $P_{dst}$ for source and destination nodes respectively,  as in Fig.~\ref{fig:square}. 
Assume now the following case holds: the distance between $P_{src}$ and its closest  replica $P^{ctr}_{c}$ is smaller than between $P_{dst}$ and its closest replica. Now the total distance  between $P_{src}$ and $P_{dst}$ is computed by summing two terms: the distance from $P_{src}$ to the closest replica $P^{ctr}_{c}$, and the one from such replica $P^{ctr}_{c}$ to $P_{dst}$.
If the considered case does not hold, the result is identical for symmetry.
Fig.~\ref{fig:avgdata} shows the average total distance $\hat{d}_{data}$ obtained by randomly generating $10^7$  pairs of nodes.
When the number of replicas is large,  $\hat{d}_{data}$ asymptotically approaches 0.5412 coherently with well-known theoretical results~\cite{gaboune1993expected}.

We now evaluate the overall synchronization traffic between the replicas, by knowing the predefined positions of the replicas in the unit square.
The average distance between any two replicas $\hat{d}_{sync}$ asymptotically approaches 0.5221 as shown in Fig.~\ref{fig:avgdata}. 
Thanks to~\eqref{eq:bb}, the synchronization traffic between the $C$ replicas can be computed as follows:
\begin{equation}\label{eq:tsync}
T_{sync} = \hat{\lambda}_{s} \hat{d}_{sync}  C(C-1) \sqrt{N}\beta
\end{equation}
where the last term considers the pair-wise synchronization between replicas. Note that $T_{sync} $ is independent from the data traffic.

Combining \eqref{eq:tdata} and \eqref{eq:tsync}, we can finally claim:
\begin{prop}
The total traffic for an unwrapped  Manhattan topology of size $N$ is given by: 
\begin{equation}\label{eq:totTraffic}
T_{TOT} = \sqrt{N}\beta({\lambda_fN\hat{d}_{data}} + {\hat{\lambda}_s\hat{d}_{sync}C(C-1)})
\end{equation}
where $\beta<\sqrt{2}$, and both $\hat{d}_{data}$ and $\hat{d}_{sync}$ depend on $C$ as shown in Fig.~\ref{fig:avgdata}.
\end{prop}

\subsection{Optimal number of replicas and its approximation}

We now evaluate numerically~\eqref{eq:totTraffic} and, through a dichotomic search, we find the optimal number of replicas that minimizes $T_{TOT}$.
Fig.~\ref{fig:NVsCopies_loglog} shows the optimal number of replicas for  different values of $N$ and $\hat{\lambda}_s/\lambda_f$.
%
\begin{figure}[tb!]
	\centering
	\includegraphics[width=8cm]{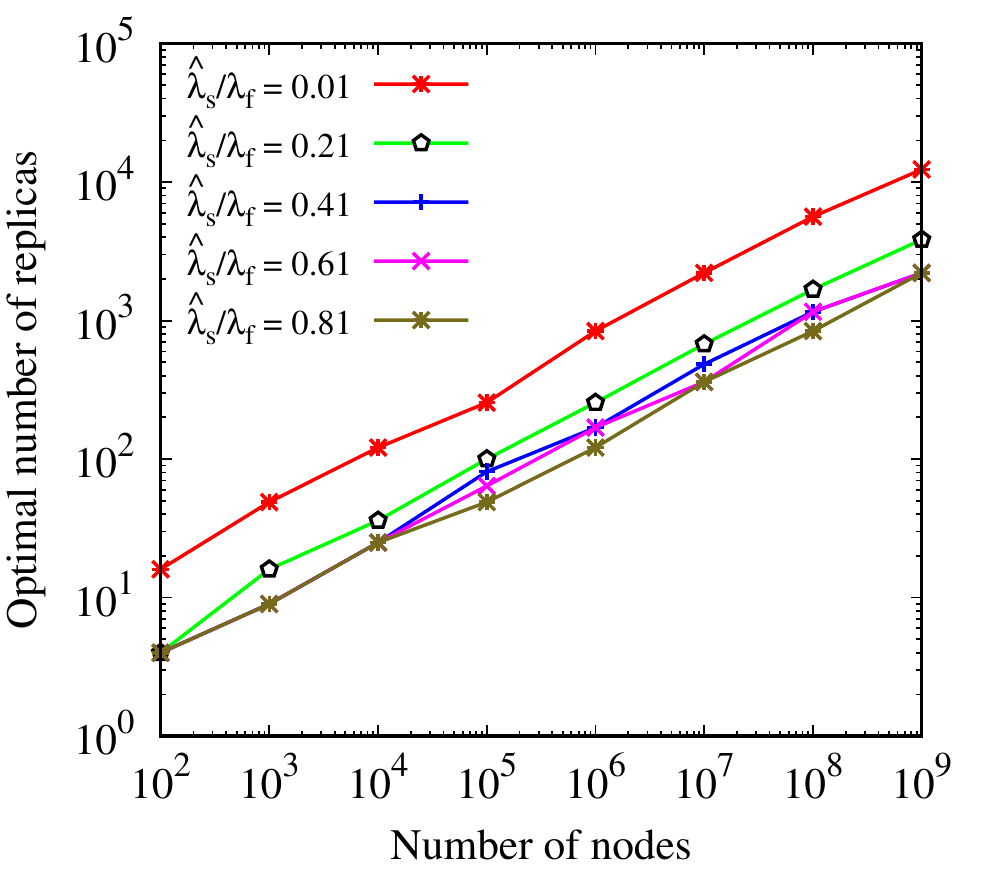}
	\caption{Optimal number of replicas for different values of $\hat{\lambda}_s/\lambda_f$.}
	\label{fig:NVsCopies_loglog}
\end{figure}
Note that for higher values of $N$, more replicas are required to cover the network. 
For higher values of $\hat{\lambda}_s/\lambda_f$, the number of replicas decreases because of the higher cost in terms of synchronization traffic. 

The curves in Fig.~\ref{fig:NVsCopies_loglog} can be fit by a function in the following form:
\begin{equation}\label{eq:curve2}
\log_{10}C_{opt}=x + y\log_{10}N+ z\log_{10} \left( \frac{\hat{\lambda}_s}{\lambda_f}\right) 
\end{equation}
with $x,y,z$ the fitting parameters.
Using standard least-square fitting procedure, we numerically evaluated the best fitting parameters and obtained the following claim:
%
\begin{prop}\label{prop2}
The optimal number of replicas $C_{opt}$ in an unwrapped  Manhattan topology of size $N$ can be approximated as follows
\begin{equation}\label{eq:curve4}
\bar{C}_{opt}=\left \lceil0.47 N^{0.40} {\left( \frac{{\lambda}_f}{\hat{\lambda}_s}\right) }^{0.40}\right \rceil
\end{equation}
which implies that $\bar{C}_{opt}$ grows as $\theta(N^{2/5})$.
\end{prop}
\begin{figure}[tb!]
	\centering
	\includegraphics[width=8cm]{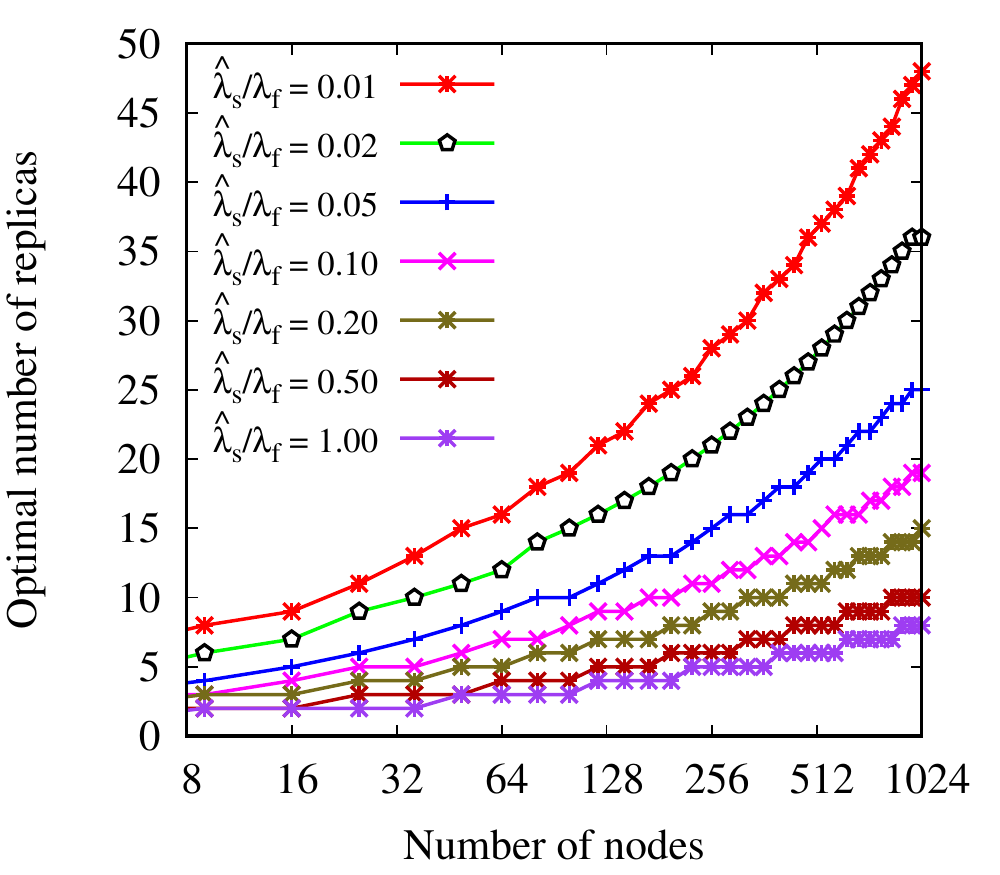}
	\caption{Optimal number of replicas  $\bar{C}_{opt}$ according to Property~\ref{prop2}.}\label{fig:prop2}
\end{figure}
Fig.~\ref{fig:prop2} shows the optimal number of replicas  $\bar{C}_{opt}$ obtained according to \eqref{eq:curve4}. As expected, if $\hat{\lambda}_s$ is small, then the number of replicas is large and for small networks correspond almost to one replica per node. For large values of \syn traffic ($\hat{\lambda}_s=\lambda_f$), the number of replicas is kept at the minimum, and 8 replicas are enough for networks with $N=1024$ switches.
We now evaluate the error introduced by Property~\ref{prop2}.
We evaluated (i) $C_{opt}$ by solving the optimization problem described in Sec.~\ref{sec:opt}, (ii) $\bar{C}_{opt}$ by computing \eqref{eq:curve4}, and (iii) the optimal number of replicas $C_{PMC}$ obtained by running PMR.
We considered the same uniform traffic pattern described in Sec.~\ref{sec:perf}  for the unwrapped Manhattan topology. All the results were obtained with 1000 different runs. 

Fig.~\ref{fig:computedVsoptimum} shows the maximum error between $\bar{C}_{opt}$ and $C_{opt}$ for $N$ that varies between $9$ and $36$. In all cases, the maximum error is bounded by one, i.e., $\bar{C}_{opt}$ overestimates by at most one the optimal number of replicas. This result shows that the formula in~\eqref{eq:curve4} is also a good approximation for small Manhattan networks. 

Due to scalability restraints we could not run the optimal solver to evaluate the error for larger networks. For this reason we had to refer to the optimal number of replicas obtained by PMR. Fig.~\ref{fig:computedVsheuristic} shows the error between $\bar{C}_{opt}$ and $C_{PMC}$ for $N$ varying between 9 and 121. Also in this case, the maximum error is bounded by one. Thus, the expression in~\eqref{eq:curve4} appears to be a reliable approximation even for larger unwrapped Manhattan topologies.

\begin{figure}[tb!]
	\centering
	\includegraphics[width=8cm]{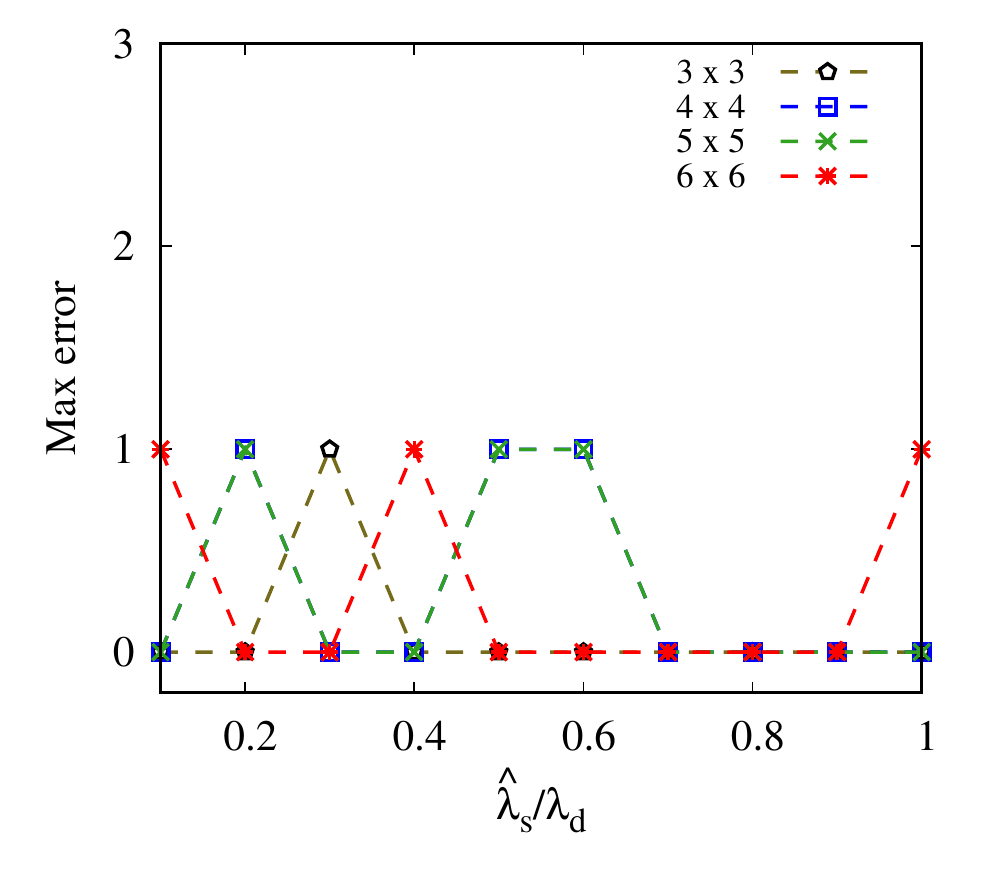}
	\caption{Maximum error $\bar{C}_{opt}-C_{opt}$ in number of replicas between the approximated formula and the optimization model.}  
	\label{fig:computedVsoptimum}
\end{figure}

\begin{figure}[tb!]
	\centering
	\includegraphics[width=8cm]{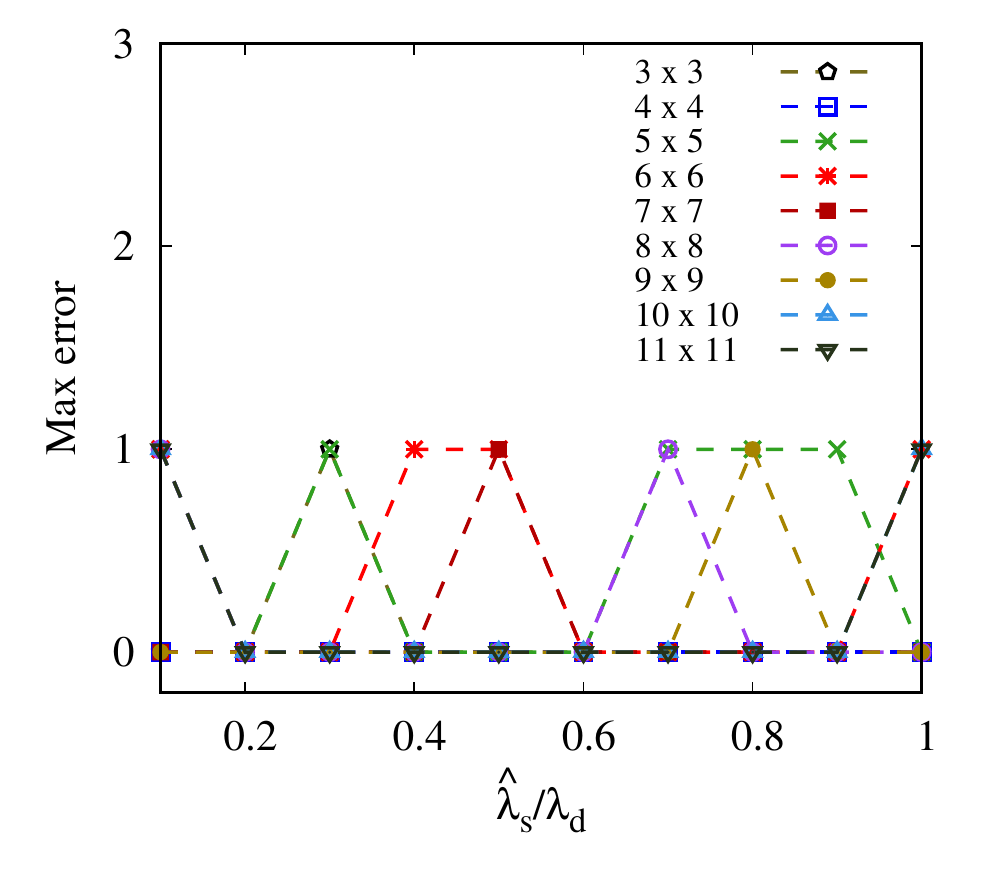}
	\caption{Maximum error $\bar{C}_{opt}-C_{PMR}$ in number of replicas between the approximated formula and the one obtained with PMR.}  
	\label{fig:computedVsheuristic}
\end{figure}

}


\section{Related works}\label{sec:statefulrelated}

{The works in~\cite{netsoft,arxiv} propose the programming abstractions to define network applications based on global states, as assumed in this work. Furthermore, they show the practical feasibility of replicating the states by describing and testing an implementation based on programmable data planes,  such as P4~\cite{bosshart2013forwarding} and Open Packet Processor (OPP)~\cite{bianchi2016open}. Both~\cite{netsoft,arxiv} assume to know the number of replicas and their placement within the network, i.e., they need an optimization engine which solves the multi-replica placement problem addressed here. On the other end, our work needs a practical implementation scheme to support the state replication as described by the two cited papers. Thus, this work and \cite{netsoft,arxiv} are complementary.}

Regarding the optimization problem addressed in this paper, the Virtual Network Embedding (VNE) problem finds the optimal placement of chains of VNFs under various optimization metrics. VNE can be closely mapped to the problem mentioned in this paper, if we consider network functions to be states and chains to be dependency graphs as computed by SNAP. Several ILP formulations and heuristics for VNE were proposed (an extensive survey is available in~\cite{fischer2013virtual}), some of which are similar to the one proposed here. However, to the best of our knowledge, none of them consider the possibility of having replicated virtual functions, the peculiar feature of this work. 

SNAP~\cite{snap16} solves the problem of the optimal placement of the states across network switches, taking into account the dependency between states and the traffic flows. However, by design, SNAP enables only one replica of each state within the network. {This limits SNAP applicability, and may impair network performance, as discussed in Sec.~\ref{sec:repl2}. To overcome this issues, we extend SNAP by enabling multiple replicas of the same state. }

 Several other network programming abstractions were proposed~\cite{kim2015kinetic, yuan2014netegg, beckett2016temporal}. However, most of them keep the states at the controller, with few existing works exploiting stateful data planes to store states.
NetKAT~\cite{mcclurg2016event} focuses on stateful data planes and provides a native support for replicated states, but, by design, the replicas are placed at the network edge (i.e., entry and exit switches) for all flows. Thus, the placement is not optimized with respect to the traffic matrix. However, our methodology could be directly applied to NetKAT.
Furthermore, the synchronization traffic is carried in piggybacking over the data traffic. Thus, both the synchronization and  the data traffic must traverse all state replicas. Instead, our proposal decouples data traffic and synchronization traffic, thus leading to more flexibility for the routing strategy. 

Swing State~\cite{luo2017swing} introduces a mechanism for state migrations entirely in the data plane but, similarly to SNAP, assumes only a single replica of a state which can be migrated across the network, on demand.

\section{Conclusions}\label{sec:statefulconc}

We consider stateful data planes, with state replication in multiple switches. We define an ILP formalization of the problem that identifies the optimal placement for the state replicas and the optimal routing for the data and \syn traffic. To cope with the limited scalability of the ILP solver, we propose the PMR algorithm and we show that it well approximates the optimal solution. 
We also numerically show the beneficial effect of state replication in the reduction of the overall traffic load in the network. Finally, we provide an  asymptotic analysis to compute the optimal number of state replicas in unwrapped Manhattan topology and show its applicability also to small graphs.
Our results advocate the adoption of replicated states when the network application is distributed and the states are ``global'' across multiple switches. Notably, our work is complementary to the works showing the feasibility of implementing replicated states in state-of-art programmable data planes. 
%

%

\bibliographystyle{IEEEtran}
\bibliography{biblio}

\begin{thebibliography}{10}
\providecommand{\url}[1]{#1}
\csname url@samestyle\endcsname
\providecommand{\newblock}{\relax}
\providecommand{\bibinfo}[2]{#2}
\providecommand{\BIBentrySTDinterwordspacing}{\spaceskip=0pt\relax}
\providecommand{\BIBentryALTinterwordstretchfactor}{4}
\providecommand{\BIBentryALTinterwordspacing}{\spaceskip=\fontdimen2\font plus
\BIBentryALTinterwordstretchfactor\fontdimen3\font minus
  \fontdimen4\font\relax}
\providecommand{\BIBforeignlanguage}[2]{{%
\expandafter\ifx\csname l@#1\endcsname\relax
\typeout{** WARNING: IEEEtran.bst: No hyphenation pattern has been}%
\typeout{** loaded for the language `#1'. Using the pattern for}%
\typeout{** the default language instead.}%
\else
\language=\csname l@#1\endcsname
\fi
#2}}
\providecommand{\BIBdecl}{\relax}
\BIBdecl

\bibitem{ref1}
D.~Kreutz, F.~M.~V. Ramos, P.~Esteves~Ver{\'i}ssimo, C.~Esteve~Rothenberg,
  S.~Azodolmolky, and S.~Uhlig, ``{S}oftware-{D}efined {N}etworking: A
  comprehensive survey,'' \emph{Proceedings of the IEEE}, vol. 103, no.~1, pp.
  14--76, Jan 2015.

\bibitem{yeganeh2013scalability}
S.~H. Yeganeh, A.~Tootoonchian, and Y.~Ganjali, ``On scalability of
  software-defined networking,'' \emph{IEEE Communications Magazine}, vol.~51,
  no.~2, pp. 136--141, 2013.

\bibitem{bosshart2013forwarding}
P.~Bosshart and al., ``Forwarding metamorphosis: Fast programmable match-action
  processing in hardware for {SDN},'' in \emph{ACM SIGCOMM CCR}, 2013.

\bibitem{bianchi2016open}
M.~Bonola, R.~Bifulco, L.~Petrucci, S.~Pontarelli, A.~Tulumello, and
  G.~Bianchi, ``Implementing advanced network functions for datacenters with
  stateful programmable data planes,'' in \emph{LANMAN}.\hskip 1em plus 0.5em
  minus 0.4em\relax IEEE, 2017, pp. 1--6.

\bibitem{7997297}
A.~Bianco, P.~Giaccone, S.~Kelki, N.~M. Campos, S.~Traverso, and T.~Zhang,
  ``{On-the-fly traffic classification and control with a stateful SDN
  approach},'' in \emph{IEEE ICC}, May 2017, pp. 1--6.

\bibitem{he2015measuring}
K.~He, J.~Khalid, A.~Gember-Jacobson, S.~Das, C.~Prakash, A.~Akella, L.~E. Li,
  and M.~Thottan, ``Measuring control plane latency in {SDN}-enabled
  switches,'' in \emph{ACM SIGCOMM SOSR}, 2015.

\bibitem{kim2015band}
\BIBentryALTinterwordspacing
C.~Kim, P.~Bhide, E.~Doe, H.~Holbrook, A.~Ghanwani, D.~Daly, M.~Hira, and
  B.~Davie, ``{In-band Network Telemetry (INT)},'' 2016. [Online]. Available:
  \url{https://p4.org/assets/INT-current-spec.pdf}
\BIBentrySTDinterwordspacing

\bibitem{netsoft}
G.~Sviridov, M.~Bonola, A.~Tulumello, P.~Giaccone, A.~Bianco, and G.~Bianchi,
  ``{LODGE: LOcal Decisions on Global statEs in programmable data planes},'' in
  \emph{IEEE NetSoft}, June 2018, pp. 257--261.

\bibitem{arxiv}
\BIBentryALTinterwordspacing
------, ``{LODGE: LOcal Decisions on Global statEs in programmable data
  planes},'' \emph{arXiv preprint arXiv:2001.07670}, 2020. [Online]. Available:
  \url{http://arxiv.org/abs/2001.07670}
\BIBentrySTDinterwordspacing

\bibitem{snap16}
M.~T. Arashloo, Y.~Koral, M.~Greenberg, J.~Rexford, and D.~Walker, ``{SNAP}:
  Stateful network-wide abstractions for packet processing,'' in \emph{ACM
  SIGCOMM}, 2016.

\bibitem{mcclurg2016event}
J.~McClurg, H.~Hojjat, N.~Foster, and P.~{\v{C}}ern{\`y}, ``Event-driven
  network programming,'' in \emph{ACM SIGPLAN Notices}, vol.~51, no.~6, 2016,
  pp. 369--385.

\bibitem{ref11}
E.~Brewer, ``{CAP twelve years later: How the ``rules'' have changed},''
  \emph{Computer}, vol.~45, no.~2, pp. 23--29, Feb. 2012.

\bibitem{paxos}
L.~Lamport, ``Paxos made simple,'' \emph{ACM Sigact News}, 2001.

\bibitem{raft}
D.~Ongaro and J.~K. Ousterhout, ``In search of an understandable consensus
  algorithm.'' in \emph{USENIX Annual Technical Conference}, 2014.

\bibitem{birman2007promise}
K.~Birman, ``The promise, and limitations, of gossip protocols,'' \emph{ACM
  SIGOPS Operating Systems Review}, vol.~41, no.~5, pp. 8--13, 2007.

\bibitem{shapiro2011conflict}
M.~Shapiro, N.~Pregui{\c{c}}a, C.~Baquero, and M.~Zawirski, ``Conflict-free
  replicated data types,'' in \emph{Symposium on Self-Stabilizing
  Systems}.\hskip 1em plus 0.5em minus 0.4em\relax Springer, 2011, pp.
  386--400.

\bibitem{petersen1996bayou}
K.~Petersen, M.~Spreitzer, D.~Terry, and M.~Theimer, ``Bayou: replicated
  database services for world-wide applications,'' in \emph{ACM SIGOPS European
  workshop}, 1996, pp. 275--280.

\bibitem{Luenberger:2015:LNP:2843008}
D.~G. Luenberger and Y.~Ye, \emph{Linear and Nonlinear Programming}.\hskip 1em
  plus 0.5em minus 0.4em\relax Springer Publishing Company, Incorporated, 2015.

\bibitem{Megiddo84}
N.~Megiddo, ``Linear programming in linear time when the dimension is fixed,''
  \emph{Journal of ACM}, vol.~31, no.~1, pp. 114--127, Jan. 1984.

\bibitem{cluster}
S.~E. Schaeffer, ``Survey: Graph clustering,'' \emph{Computer Science Review},
  vol.~1, no.~1, pp. 27--64, Aug. 2007.

\bibitem{voronoi}
K.~Ruddel and A.~Raith, ``Graph partitioning for network problems,'' in
  \emph{Joint NZSA ORSNZ Conference}, no. 107, 2013, pp. 1--10.

\bibitem{ibmcplex}
\BIBentryALTinterwordspacing
``{CPLEX Optimizer}.'' [Online]. Available:
  \url{https://www.ibm.com/analytics/cplex-optimizer}
\BIBentrySTDinterwordspacing

\bibitem{watts1998collective}
D.~J. Watts and S.~H. Strogatz, ``Collective dynamics of 'small-world'
  networks,'' \emph{Nature}, vol. 393, no. 6684, p. 440, 1998.

\bibitem{gaboune1993expected}
B.~Gaboune, G.~Laporte, and F.~Soumis, ``Expected distances between two
  uniformly distributed random points in rectangles and rectangular
  parallelpipeds,'' \emph{Journal of the Operational Research Society},
  vol.~44, no.~5, pp. 513--519, 1993.

\bibitem{fischer2013virtual}
A.~Fischer, J.~F. Botero, M.~T. Beck, H.~De~Meer, and X.~Hesselbach, ``Virtual
  network embedding: A survey,'' \emph{IEEE Communications Surveys \&
  Tutorials}, vol.~15, no.~4, pp. 1888--1906, 2013.

\bibitem{kim2015kinetic}
H.~Kim, J.~Reich, A.~Gupta, M.~Shahbaz, N.~Feamster, and R.~Clark, ``Kinetic:
  Verifiable dynamic network control,'' in \emph{USENIX NSDI 15}, 2015, pp.
  59--72.

\bibitem{yuan2014netegg}
Y.~Yuan, R.~Alur, and B.~T. Loo, ``{NetEgg}: Programming network policies by
  examples,'' in \emph{ACM SIGCOMM HotNets}, 2014, p.~20.

\bibitem{beckett2016temporal}
R.~Beckett, M.~Greenberg, and D.~Walker, ``Temporal {NetKAT},'' \emph{ACM
  SIGPLAN Notices}, vol.~51, no.~6, pp. 386--401, 2016.

\bibitem{luo2017swing}
S.~Luo, H.~Yu, and L.~Vanbever, ``{Swing State}: Consistent updates for
  stateful and programmable data planes,'' in \emph{ACM SIGCOMM SOSR}, 2017.

\end{thebibliography}

\end{document}